\documentclass[twocolumn]{bmcart}

\usepackage{amsthm,amsmath, amssymb}
\RequirePackage[authoryear]{natbib}
\usepackage[utf8]{inputenc} 
\usepackage{color}
\usepackage{comment}
\usepackage{graphicx}
\usepackage{transparent}
\DeclareMathOperator\erf{erf}

\startlocaldefs
\endlocaldefs

\begin{document}

\begin{frontmatter}

\begin{fmbox}
\dochead{Research}


\title{A New Hybrid Technique for Modeling Dense Star Clusters}


\author[
   addressref={aff1},                   
   corref={aff1},                       
   noteref={n1},                        
   email={carlrodr@mit.edu}   
]{\inits{CL}\fnm{Carl L} \snm{Rodriguez}}
\author[
   addressref={aff2,aff3,aff4},
]{\inits{B}\fnm{Bharath} \snm{Pattabiraman}}
\author[
   addressref={aff2,aff3},
]{\inits{S}\fnm{Sourav} \snm{Chatterjee}}
\author[
   addressref={aff2,aff4},
]{\inits{A}\fnm{Alok} \snm{Choudhary}}
\author[
   addressref={aff2,aff4},
]{\inits{W}\fnm{Wei-keng} \snm{Liao}}
\author[
   addressref={aff2},
]{\inits{M}\fnm{Meagan} \snm{Morscher}}
\author[
   addressref={aff2},
]{\inits{M}\fnm{Frederic A} \snm{Rasio}}


\address[id=aff1]{
  \orgname{MIT-Kavli Institute for Astrophysics and Space Research}, 
  \street{77 Massachusetts Ave, 37-664H},                     %
  \postcode{02139}                                
  \city{Cambridge, MA},                              
  \cny{USA}                                    
}
\address[id=aff2]{%
  \orgname{Center for Interdisciplinary Exploration and Research in
  37 Astrophysics (CIERA) },
  \street{2145 Sheridan Rd},
  \postcode{60208}
  \city{Evanston, IL},
  \cny{USA}
}
\address[id=aff3]{%
  \orgname{Dept.~of Physics and Astronomy, Northwestern University},
  \street{2145 Sheridan Rd},
  \postcode{60208}
  \city{Evanston, IL},
  \cny{USA}
}
\address[id=aff4]{%
  \orgname{Dept.~of Electrical Engineering and Computer Science, Northwestern University},
  \street{2145 Sheridan Rd},
  \postcode{60208}
  \city{Evanston, IL},
  \cny{USA}
}


\begin{artnotes}
\note[id=n1]{Pappalardo Fellow} 
\end{artnotes}



\begin{abstractbox}

\begin{abstract} 

The ``gravitational million-body problem,''  to model the dynamical evolution of
a self-gravitating, collisional $N$-body system with $\sim 10^6$ particles over many
relaxation times, remains a major challenge in computational astrophysics.
Unfortunately, current techniques to model such systems suffer from severe
limitations. A direct $N$-body simulation with more than $10^5$ particles can
require months or even years to complete, while an orbit-sampling Monte Carlo
approach cannot adequately model the dynamics in a dense cluster core, particularly
in the presence of many black holes. We have developed a new technique combining
the precision of a direct $N$-body integration with the speed of a Monte Carlo approach.
Our Rapid And Precisely Integrated Dynamics code, the \texttt{RAPID} code,
statistically models interactions between neighboring stars and stellar binaries
while integrating directly the orbits of stars or black holes in the cluster core. This allows
us to accurately simulate the dynamics of the black holes in a realistic
globular cluster environment without the burdensome $N^2$ scaling of a full
$N$-body integration. We compare \texttt{RAPID} models of idealized globular
clusters to identical models from the direct $N$-body and
Monte Carlo methods. Our tests show that
\texttt{RAPID} can reproduce the half-mass radii, core radii,
black hole ejection rates, and binary properties of the
direct $N$-body models far more accurately than a standard Monte Carlo
integration while remaining significantly faster than a
full $N$-body integration. With this technique,
it will be possible to create more realistic models of Milky Way globular clusters with sufficient rapidity to explore the full parameter space of dense stellar clusters.

\end{abstract}




\end{abstractbox}
\end{fmbox}

\end{frontmatter}




\section{Main Text}

The dynamics of dense star clusters is one of the most challenging
problems of modern computational astrophysics.  The large number of
particles, high interaction rate, and large number of processes with vastly different physical timescales conspire to make globular clusters (GCs) and galactic
nuclei (GN) uniquely difficult to model.  In particular, the large number of black holes (BHs) in both GCs and GN
often dynamically interact on a much shorter timescales than the rest of the cluster \citep{Spitzer1969}.  Although only comprising a small fraction of the
total cluster mass, these BHs provide the dominant energy source for GCs, especially after the BH-driven core-collapse \citep{Morscher2015}.  Hence, understanding their dynamics is critical to understanding the overall evolution and present-day appearance of these systems \citep{Mackey2008}.  Unfortunately, since the orbital and interaction timescales of these BHs are frequently orders-of-magnitude smaller than the interaction timescale of a typical star in the cluster, resolving these effects can be particularly difficult.

Modern stellar dynamics codes have intensely investigated GCs, with the majority
of work focusing on two approaches.  The $N$-body approach directly integrates
the force of every particle on every other particle, with the current generation
of codes
\citep{PortegiesZwart2001a,Harfst2008,Nitadori2012,Capuzzo-Dolcetta2013a,Wang2015}
making extensive use of state-of-the-art hardware acceleration and algorithmic
enhancements.  While extremely precise, this approach can require more than a year
\citep[e.g.,][]{Heggie2014b,Wang2016} to complete a full simulation of a realistic
Milky-Way GC.

As such, an approximate Monte Carlo (MC) technique is often used in place of a
full direct summation \citep{Henon1971,Giersz1998,Joshi1999,Freitag2001}.
Whereas an $N$-body approach computes the orbit of stars directly, the
orbit-sampling approach assumes that particle orbits remain fixed on a dynamical
timescale, only changing due to slight perturbations from two-body encounters
between neighboring particles.  This allows the orbits to be sampled
statistically, and since computing a single orbit in a fixed spherical potential
is faster than computing the precise orbits in the full-$N$
potential of a cluster, these MC models can be generated in at most a few days
or weeks.  However, the assumptions of spherical symmetry and dynamical
equilibrium break down in the BH-dominated core, where the potential and the
particle orbits are primarily determined by a small number of particles.  This 
can lead to a substantial underprediction of the core radii by MC techniques (compared to direct $N$-body), particularly during the deep collapses that produce dynamically-assembled binaries \citep{Morscher2015,Rodriguez2016a}.

GCs are formed as the result of a burst of star formation in the early universe.
Approximately 10 to 20 Myr after this formation is complete, the most massive stars in the cluster collapse, yielding hundreds to thousands of BHs
\citep{Belczynski2006}.  As the BHs are
more massive than the typical cluster star, they are rapidly driven to the center of
the GC by dynamical friction \citep{Fregeau2002}; once there, the number density 
of BHs is sufficient to form binaries via three-body encounters.  While it was 
long-assumed that these BHs would not be retained in GCs to the present day 
\citep[e.g.,][]{Sigurdsson1993}, recent evidence has begun to suggests otherwise.

The past decade has seen the first detections of BHs in GCs, starting with the first detection in an
extragalactic GC by \cite{Maccarone2007a} and several recent detections in
Milky Way GCs, \citep{Chomiuk2013,Strader2013,Miller-Jones2015}, including two BH candidates in M22 \citep{Strader2012} and the recent dynamical measurement of a $\gtrsim 4.5 M_{\odot}$ BH in NGC 3201 \citep{2018MNRAS.475L..15G}.  {These observational results complimented recent theoretical results suggesting {
that the GCs can potentially retain hundreds of BHs up to the present day 
\citep{Mackey2007,Downing2012,Morscher2012,Morscher2015,Kremer2018,Askar2018}.  
This has led to a new theoretical understanding that the number of BHs retained 
in a GC directly controls the size and density of its observational core 
\citep[e.g.,][]{Merritt2004,Mackey2008,Sippel2013,Breen2013,Kremer2018a,ArcaSedda2018}}  }The importance of BHs in GCs cannot be overstated.  In addition to determining the structural and evolutionary properties of the clusters, GCs also have important implications for BH astrophysics.
GCs can produce X-ray binaries at a significantly higher rate than the galactic
field \citep{Clark1975X-rayClusters}, suggesting that there might be $\sim 100$s
of low-mass X-ray binaries in Milky Way GCs \citep{Pooley2003}.  Furthermore,
recent studies have shown that the second-generation of gravitational-wave
detectors can potentially detect $\gtrsim 100$ binary BH mergers per year from
binaries forged in the cores of GCs
\citep{Rodriguez2015a,Antonini2015,Rodriguez2016b}, with recent detections \citep[e.g.~GW170104, ][]{2017PhRvL.118v1101A} by LIGO/Virgo showing spin alignments suggestive of dynamical formation. As such, understanding the dynamics of these systems is critical.

What is needed is a technique that combines the speed of the MC approach with
the precision of a direct $N$-body integration.  In this paper, we describe a
new code, the \emph{Rapid and Precisely Integrated Dynamics} (\texttt{RAPID})
code, which combines both methods into a ``best of both worlds'' approach.  In
this method, the majority of particles are modeled with our parallel
H\'enon-style code, the Cluster MC (\texttt{CMC}) code \citep{Pattabiraman2013},
while the orbits of BHs are integrated directly with the Kira $N$-body
integrator \citep{PortegiesZwart2001a}.  We find that this technique accurately
reproduces the core radii and BH dynamics of a full direct $N$-body integration,
with similar runtimes to the MC approach.  Although we only integrate the BH
orbits directly in the current work, the method is general, allowing us to
select any population of particles in the cluster for $N$-body integration.

In Section 2, we briefly review the $N$-body and MC approaches, and describe the
combination of the two approaches as implemented in \texttt{RAPID} code.  In
Section 3, we describe a single \texttt{RAPID} timestep, illustrating the
technical details of the approach, while in Section 4, we describe the
parallelization strategy that allows us to compute particle positions and
velocities via orbit sampling and direct $N$-body simultaneously.  {In Section 5,
we show the results of an analytic toy model, comparing the inspiral due to dynamical friction of a single particle as
predicted by theory, direct $N$-body, and \texttt{RAPID}}.  Finally, in Section 6, we compare the properties of four idealized GCs as modeled by \texttt{NBODY6}, \texttt{CMC}, and \texttt{RAPID}.  Throughout the paper, we will frequently refer to the ``stars'' and ``BHs'' in the cluster separately.  In our current method, the stars are modeled with CMC and the BHs are integrated with Kira.  This shorthand is to delineate which systems are being modeled by which technique, even though the particles under consideration are point-mass particles.

\section{Hybridization Approach}

In this section, we provide a brief overview of the current methods employed to model GCs, and describe how our approach combines the virtues of both methods.  Both the $N$-body and MC approaches are the result of decades of precision work by multiple groups.  For a more comprehensive description of collisional $N$-body dynamics, see \cite{Aarseth2003} or \cite{Dehnen2011}.  A review of MC methods can be found in \cite{Freitag2008}.

It should be noted that \texttt{RAPID} is not the first attempt at a hybrid
$N$-body/statistical sampling approach to stellar dynamics.  In particular, the
hybrid approach developed by \cite{McMillan1984} combined a Fokker-Planck
sampling code with a direct $N$-body approach, in order to study GCs undergoing
core collapse \citep{McMillan1984a,McMillan1986}.  The \texttt{RAPID} code
continues this tradition of attempting to ``have it all'', by combining the best of the direct integration and statistical sampling methods.

\begin{figure*}
\centering
\includegraphics[trim=0cm 2cm 0cm 2cm,width=0.95\textwidth]{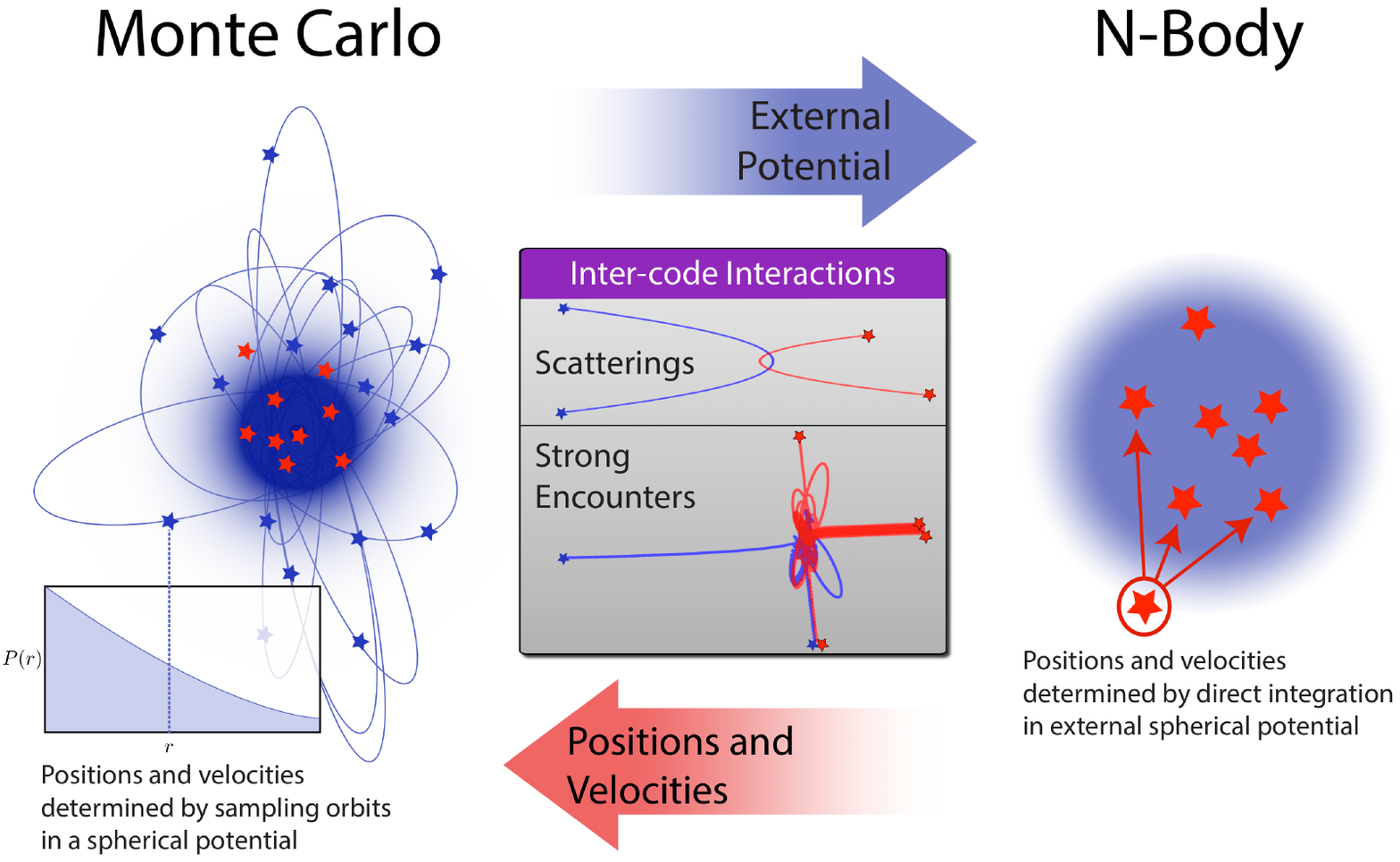}
\caption{\textbf{Cartoon explanation of the \texttt{RAPID} Code}  The majority of particles are evolved via the MC approach, with their positions and velocities determined by randomly sampling orbits in the spherical potential of the entire cluster.  Meanwhile, a small handful of the most massive objects (the BHs) have their positions and velocities determined by direct $N$-body integration inside the external potential of the MC stars.  Particles from both codes can interact via two-body encounters and strong scattering encounters, allowing energy to flow between the two systems.}
\label{fig:flowchart_Kira}
\end{figure*}

\subsection{Direct \textit{N}-Body Integration}

The physical principle behind a direct
$N$-body integrator is simple: since the force on any given particle in the sum
of the gravitational force from every other particles in a given system, the most
accurate way to model such a system is to numerically sum all the forces.  This
is the underlying principle behind the $N$-body integrators.  The most
frequently used of these codes, the \texttt{NBODY} series of codes, have been improved and finely tuned with additional physics, including stellar evolution \citep{Hurley2001}, algorithmic regularization \citep{Aarseth1999}, post-Newtonian chain regularization, \citep{Aarseth2012}, and GPU acceleration \citep{Nitadori2012}.  With advanced hardware and a minimal number of simplifying assumptions, direct integration is the most precise method available for modeling dense stellar systems.

However, this precision comes at a cost.  {Naively, the cost of an 
$N$-body integration scales as $N^2$, since one must evaluate the force of every 
particle on every other particle, every timestep.  In practice, most modern 
$N$-body codes do not evaluate the force between every particle every timestep, 
instead opting for a variable ``block'' timestep approach in which only certain 
particles have their forces re-evaluated at a given time.  Despite this, and 
many other algorithmic improvements (such as employing a nearest neighbor scheme to 
accelerate force evaluations), the computational cost to integrate a cluster of 
$N$ particles forward by a given physical time scales as $\mathcal{O}(N^2)$, regardless of the timestep scheme or mass distribution of the cluster \citep{Makino1988}}.

{This steep scaling makes large-scale simulations of massive star clusters exceedingly challenging.}  The largest simulation attempted
with \texttt{NBODY6} is currently the $N = 5\times10^5$ model of galactic GC M4
 performed by \cite{Heggie2014b}, requiring 2.5 years on a dedicated GPU system.  More recently, the
current state-of-the-art parallelized code \texttt{NBODY6++GPU} \citep{Wang2015} can model a realistic ($N = 10^6$) cluster in little more than a
year \citep{Wang2016}.  Despite these remarkable achievements, simulation times in excess of $\sim1$ year for large
systems preclude any reasonable exploration of the parameter space of initial
conditions of GCs, and any collisional models of GN ($N = 10^7 - 10^9$) remain beyond the capabilities of the
current generation of direct summation techniques.  To answer astrophysical
questions related to such systems, a more rapid technique is called for.

For our hybrid approach, we use the Kira $N$-body integrator, included as part
of the \texttt{Starlab} software package \citep{PortegiesZwart2001a}.  Like the
\texttt{NBODY} series of codes, Kira is a $4^{th}$-order Hermite
predictor-corrector integrator with a block timestep scheme.  Kira also integrates close encounters and tightly-bound multiples using Keplerian regularization, where sufficiently-isolated hyperbolic and tightly bound binaries are evolved as analytic two-body systems.  Additionally, Kira organizes its internal data using easily-modifiable C++ class structures, and includes an easily-customizable module for including an external gravitational potential.  These two features make it ideal for inclusion in the hybrid method.

\subsection{Orbit-Sampled Monte Carlo}

The MC approach assumes that the large-scale
evolution of a star cluster can be modeled as slow transitions
from one equilibrium configuration to the next.  These transitions are driven by two-body scatterings (relaxations) between particles in the cluster.  Because it is the statistically-predictable effect of many of these two-body scatterings that let energy flow through the cluster, one can describe the cumulative effect of these relaxations as a single, effective two-body encounter.  We often describe
these systems in terms of their dynamical timescales (the crossing time for a
single particle) and their relaxation timescales (the average time for the
velocity of a single particle to change by a certain amount). For a system in dynamical equilibrium:

\begin{equation}
T_{\text{relax}} \approx \frac{0.1 N}{\log N} T_{\text{dyn}} \gg T_{\text{dyn}}
\end{equation}

 \noindent where $N$ is the number of particles \citep[see][]{Binney2008}.  With a
sufficiently large $N$, a star cluster can be thought of as a series of
independent orbits that only change on a $T_{\text{relax}}$ timescale.
This assumption eliminates the need to directly compute the
forces upon a single particle on an orbital timescale.  Instead, we only need to
determine the shape of a star's orbit after it has changed velocity due to
encounters with other stars on the relaxation timescale.

For some cases. such as a spherically asymmetric mass distribution, the orbit of the particle must be integrated numerically
\citep[e.g.,][]{Vasiliev2014a,Vasiliev2015}; however, for most
applications to large collisional star clusters (such as GCs and GN) the background gravitational potential can be assumed to be
spherical.  This allows the clever theorist to determine a star's position and velocity by analytically
sampling a random point along its orbit.
This \emph{orbit-sampling MC} approach, first developed by \cite{Henon1971} and built upon by multiple groups \citep{Stodoikiewicz1982,Giersz1998,Joshi1999,Freitag2001} can model stellar systems with $N \gtrsim 10^7$ particles in a fraction of
the time of a direct $N$-body simulation.  {Unlike a direct $N$-body integration, the orbit calculation and dynamical encounters in the MC method scale linearly with the number of particles; only the sorting of particles by radius, with its characteristic $N\log N$ complexity, limits the scaling.}  {Furthermore, the MC method computes the interactions of particles on a relaxation timescale, as opposed to the dynamical timescale of a direct $N$-body integration.}  Put together, the computational difficulty of the MC method scales as $\mathcal{O}(N\log N)$ per half-mass relaxation time, versus $\mathcal{O}(N^3)$ for a direct $N$-body approach.  Because of this, the
MC method can easily model large systems that are simply beyond the
reach of other techniques.

However, the assumptions that enable the speed of the MC method can
easily break down in some of the most interesting regions of parameter space.
Once mass segregation is complete, the evolution of a GC is largely determined
by the small number of BHs that have accumulated in the core.  This can
consist of as few as hundreds or even tens of BHs.  Since the dynamics of these
small, spherically asymmetric systems change rapidly on an orbital timescale, the MC method is
unable to accurately follow the evolution of these BHs in the center of the
cluster.  And since this small cluster of BHs forms the hard binaries whose
binding energy acts as a power source for the entire cluster, their dynamics
\emph{must} be accurately modeled to understand the long-term evolution of the cluster.

Our orbit-sampling Cluster MC code, \texttt{CMC}, was first developed by \cite{Joshi1999}, based on the original developments by \cite{Henon1971} and \cite{Stodoikiewicz1982}.  As the code considers interactions between individual stars, \texttt{CMC} incorporates multiple physical processes, including stellar evolution \citep{Hurley2000,Hurley2002}, strong three-body and four-body scatterings with the small-$N$ integrator \texttt{Fewbody} \citep{Fregeau2004}, probabilistic three-body binary formation \citep{Morscher2012}, and physical collisions.  Additionally, \texttt{CMC} has recently been parallelized to run on an arbitrary number of computer processors \citep{Pattabiraman2013}.  This MPI parallelizaion makes \texttt{CMC} an ideal code base for \texttt{RAPID}, as the current parallelization scheme can be easily expanded to allow the $N$-body integration to run in parallel to the MC

\subsection{Hybrid Partitioning}
\label{subsec:partition}

The cornerstone of any hybrid modeling technique is the domain decomposition between methods.  Since it is the BHs that are driving the non-spherical, non-equilibrium dynamics in the cluster core, the natural division is for Kira to integrate the BHs while \texttt{CMC} integrates all the remaining stars in the cluster.  By default, \texttt{RAPID} divides the system using one of two criteria:

\begin{itemize}

\item a \textbf{mass criterion}, which divides the system
according to a specified threshold, where particles above the threshold are considered BHs and particles below it are considered stars, and

\item a \textbf{stellar evolution criterion}, in which objects
identified as BHs by stellar evolution are integrated by Kira, and all other objects are integrated by \texttt{CMC}

\end{itemize}

\noindent By default, \texttt{\texttt{RAPID}} employs the first criterion for point-mass
simulations (with a user-specified threshold mass), and the second criterion for
simulations using stellar evolution.  Any mixed objects (e.g., a BH-star binary) are evolved in \texttt{CMC}, in order to treat the binary stellar evolution consistently.

{There are two reasons to focus on BHs in our hybridization scheme.  The first 
is that by limiting the integration to a persistent set of particles, we can 
avoid the the large communications overhead that is incurred each time particle must be transferred back and forth from MC to $N$-body.  This would occur much more frequently if, for instance, we divided our computational domains according to radius, with the $N$-body integrating particles in the core, and the MC integrating particles in the halo \citep[similar to][]{McMillan1984}.  Secondly, by limiting the $N$-body to only BHs, we sidestep the difficulties of treating binary stellar evolution during the $N$-body integration.  Although Kira includes a built-in package for binary and single stellar evolution (the \texttt{SeBa} package), it is not compatible with the stellar evolution in \texttt{CMC} \citep[the Binary Stellar Evolution of][]{Hurley2002}.  We will explore ways to integrate self-consistent stellar evolution into the hybrid approach in a future work.}

\begin{figure}
\centering
\includegraphics[trim=0cm 0cm 0cm 0cm,width=0.45\textwidth]{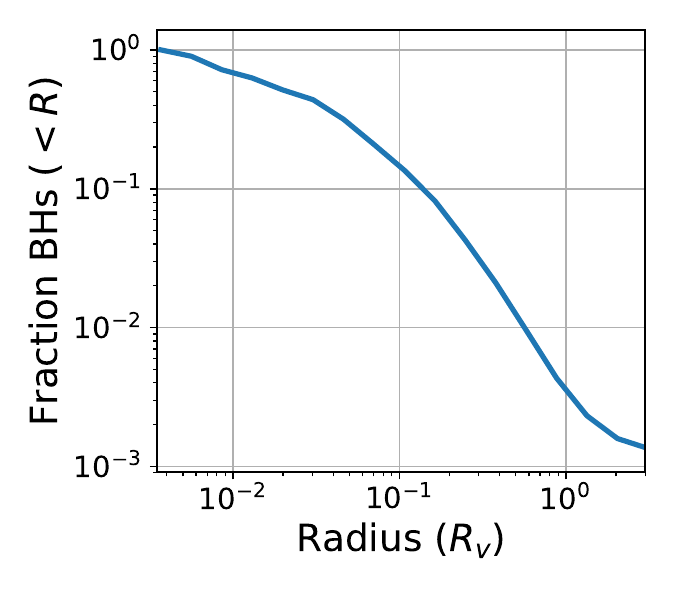}
\caption{\textbf{Mass Segregation of BHs in a Realistic GC --} The cumulative fraction of all objects that are BHs as a function of radius, taken from a realistic GC model with $N=10^6$ initial particles and realistic stellar evolution after 100 Myr of evolution \cite{Rodriguez2018}.  At the central region of the core ($r < 0.01$ $R_v$) the evolution is almost entirely dominated by the BHs, with 75\% of all objects in that region being BHs.}
\label{fig:flowchart_Kira}
\end{figure}

{However, in realistic clusters, we find that the segregation between BHs and 
stars is extreme, with the inner-most regions of the cluster completely 
dominated by BHs.  In Figure 2, we show the cumulative fraction of BHs as a 
function of cluster radius for a typical GC model with $N=10^6$ and full stellar 
evolution \cite[see][]{Rodriguez2018}.  After 100Myr, the central region of the 
cluster is completely dominated by BH, with 75\% of the objects less that 0.01 
pc from the cluster center being BHs.  These are the objects that primarily 
participate in the dynamical formation of binaries that drive the cluster 
evolution \citep{Breen2013,Morscher2015}.  {Furthermore, any 
non-spherical effects that arise from having a small number of particles in the 
cluster center will be limited to these central BHs, ensuring that the hybrid 
approach can correctly integrate the correct 3D potential in the central 
regions.}}

The \texttt{RAPID} code builds upon the \texttt{CMC} parallelization described in \cite{Pattabiraman2013}, and is designed to be run on a distributed computational system with at least 2 parallel MPI processes: one for the MC integration, and one for the $N$-body integration.
An initial \texttt{RAPID} run begins with all objects integrated with CMC on all available processes.  After a user-specified criterion is met, a single MPI process initializes the Kira $N$-body integrator.  At this point, any stars that are on that \texttt{CMC} process are transferred to the remaining \texttt{CMC} process(es), and the BHs are collected on the Kira process for the $N$-body integration.  Once both integrators have been
initialized, particles can be transferred back and forth between the Kira and
\texttt{CMC} processes (e.g., a star becomes a BH through stellar evolution, and is copied from \texttt{CMC} to Kira).  Additionally,
information about the particles must frequently be communicated back and forth between \texttt{CMC} and Kira.  At each timestep, \texttt{CMC} needs to know the positions and velocities of the BHs,
while Kira needs to know the external potential of the stars.  See Section \ref{sec:parallelization} for
the details of the parallelization strategy.
At the end of each timestep, the information from Kira and \texttt{CMC} is combined and printed to file, in order to create a coherent cluster model.  See Figure 1.

\section{\texttt{RAPID} Timestep}
\label{sec:timestep}

At its core, the main difference between \texttt{RAPID} and \texttt{CMC} is the computation of
the orbits and positions of the BHs.  Here, we describe in detail a single
\texttt{RAPID} timestep, highlighting the differences between the hybrid approach and a
standard \texttt{CMC} integration.  See Figure 3.

\begin{figure*}
\centering
\includegraphics[trim=0cm 0.cm 0cm 0cm,scale=0.45]{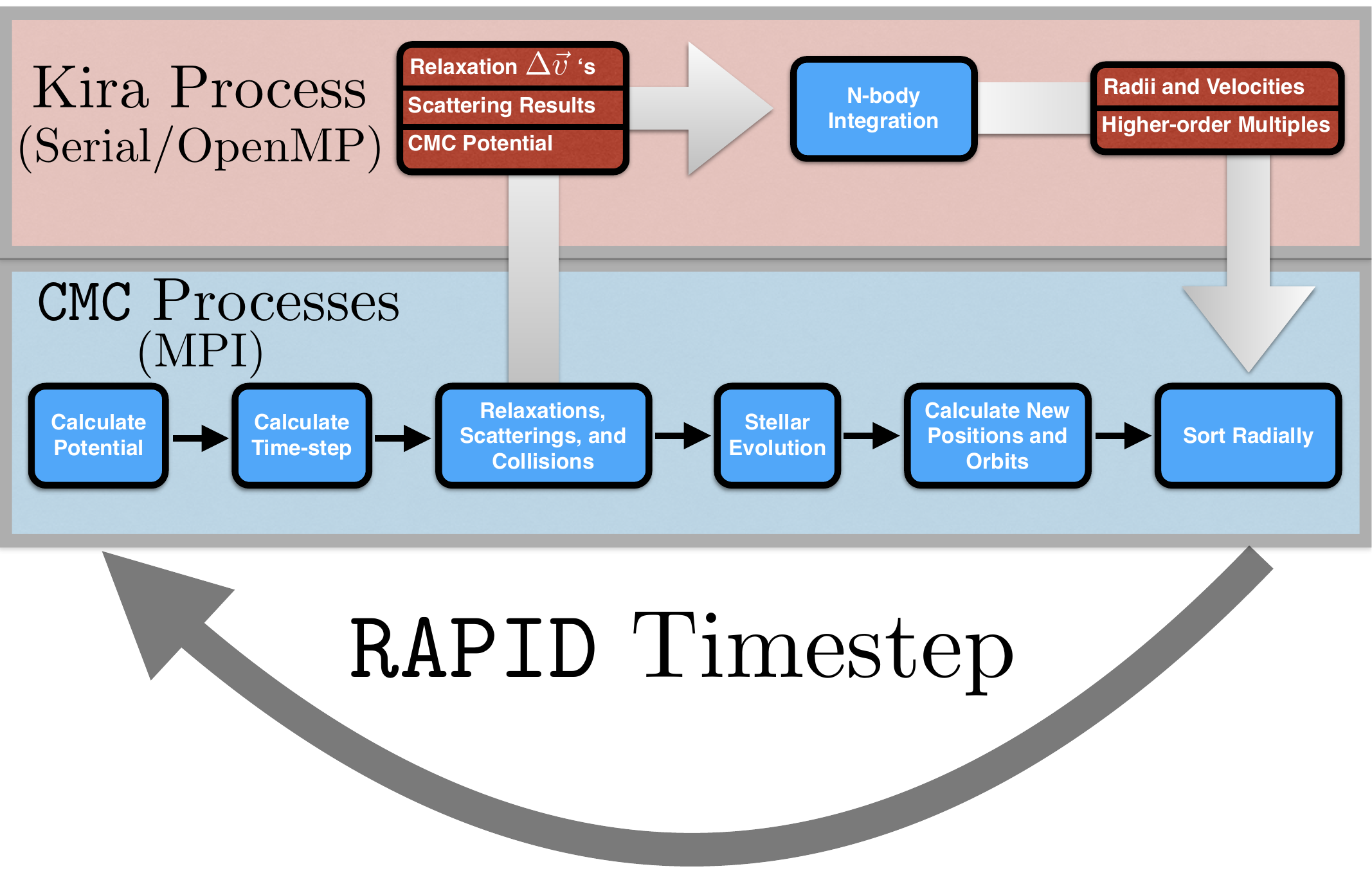}
\caption{\textbf{Flowchart of the \texttt{RAPID} Code} --  After the hybridization is
initialized, the BHs are transferred to the Kira process, while the remaining
Monte Carlo processes sample the orbital positions of the remaining stars.  The
communication between the Kira process and the \texttt{CMC} process(es) is
handled by an extension of the MPI scheme developed in \texttt{CMC} \citep{Pattabiraman2013}, while the $N$-body process can be run in serial or with OpenMP threading.}
\label{fig:timestep}
\end{figure*}

\subsection{Compute the Potential}
\label{subsec:potential}

Since the cluster is assumed to be
spherically symmetric, the gravitational potential at radius $r$  between stars
$k$ and $k+1$ can be expressed as:

 \begin{equation}
 \Phi(r) = G\left(-\frac{M_k}{r} - \sum^{N}_{i = k +1}\frac{m_i}{r_i}\right)
 \end{equation}

 \noindent where $M_j \equiv \sum^{j}_{i=1}m_i$.  As the stars are
 radially sorted, the potential at each star can be computed recursively with a  single pass from the outermost star inwards:

 \begin{align}
 	&\Phi_{N+1} = 0 \nonumber \\
 	&M_N = \sum^{N}_{i=1}m_i \nonumber \\
 	&\Phi_k = \Phi_{k+1} - G M_k\left(\frac{1}{r_k} - \frac{1}{r_{k+1}}\right) \nonumber \\
    &M_{k - 1} = M_{k} - m_k
 \end{align}
 {
 \noindent where $N$ is the number of particles, $m_i$ and $r_i$ are the mass and radius of particle $i$, $M_i$ is the total mass of particles interior to and including particle $i$.  Note that the outer boundary conditions at $N+1$ are not associated with physical particles, but are the outer boundary points of the cluster (with $r_{N+1} = \infty$). }

In the \texttt{RAPID} code, two potentials are calculated: the full spherical potential, $\Phi$,
of all stars and BHs, and a MC-only potential,
$\Phi^{\text{MC}}$, computed only with the stars in \texttt{CMC}.
The MC potential is sent from the \texttt{CMC} processes to the $N$-body process, and used as an
external potential for the Kira integration.


\subsection{Select the Timestep}
\label{subsec:timestep}

Since \texttt{CMC} and \texttt{RAPID} consider multiple
physical processes, the timestep selection must consider multiple relevant
timescales: the timescale for two-body relaxation ($T_{\text{rel}}$), the
timescales for binary-single and binary-binary strong scatterings
($T_{\text{BS}}$ and $T_{\text{BB}}$), the timescale for physical collisions
($T_{\text{coll}}$), the timescale for stellar evolution to
contribute a change in an objects mass ($T_{\text{SE}}$), and the timescale for tidal stripping of stars to alter the cluster mass ($T_{\text{tid}}$).  The
simulation timestep is then selected to be the minimum of each of these
timescales, or

\begin{equation} \Delta T_{\text{CMC}} =
\min(T_{\text{rel}},T_{\text{BS}},T_{\text{BB}},T_{\text{coll}},T_{\text{SE}},T_{\text{tid}})
\end{equation}

The  \texttt{RAPID} timestep is chosen in a similar fashion.  The timescale for each
physical process is computed for all particles in the cluster (stars and BHs),
and the minimum (or a fraction of the minimum) is selected as the current timestep.  In Kira, this timestep determines how many dynamical times the $N$-body system will be advanced.

Unlike \texttt{CMC}, \texttt{RAPID} can produce BH multiples with an arbitrary number of components.  To compute the interaction timescale for scatterings between stars and BH multiple systems,
the timestep is
chosen using the same prescription as the $T_{\text{BS}}$ and $T_{\text{BB}}$ timesteps, but with the semi-major axis of the outermost binary pair as the effective width of the system.

{The relaxation timestep, $T_{\rm{rel}}$, is selected as the minimum of }

\begin{equation}
T_{\rm{rel}} =
\frac{\theta_{\rm{max}}}{\pi/2}\frac{\pi}{32}\frac{v^3_{\rm{rel}}}{\log(\gamma
N)G^2 n (m_1+m_2)^2}
\label{eqn:trel}
\end{equation}

\noindent {for all pairs of neighboring particles \citep{Freitag2001}, where $v_{\rm{rel}}$ is the relative velocity of the two stars, $n$ is the local number density of stars.   $\theta_{\rm{max}}$ is the maximum-allowed scattering angle for
two-body relaxation.  {Quantities which are taken as local averages (such as the number density, the velocity dispersion, etc.) are all computed using an averaging kernel of 40 particles.  In other words, to compute the density of stars around star $r_i$, we average the density over all particles from $r_{i-20}$ to $r_{i+20}$.}

For the standard definition of the relaxation time from
\cite{Binney2008}, $\theta_{\rm{max}} = \pi/2$.  {However, for  the
systems considered here (particularly the highly-idealized two-component models
presented in Section \ref{sec:results}), this averaging can sometimes smooth out
the otherwise short relaxation times between a heavy object and a neighboring
lighter object (particularly if there is only one heavy object in that bin
of 40 particles).  For the theoretical comparison shown in Section
\ref{sec:plummer}, we still set $\theta_{\rm{max}}$ to the theoretical value of $\pi/2$, but average the above
quantities over the closest 2 particles.  For the numerical comparison in
Section \ref{sec:results}, we average over the nearest 40 particles, but use $\theta_{\rm{max}}
= 1$ to calculate the relaxation timestep, which was found \citep{Fregeau2007}
to provide a good compromise between accuracy and speed for such two-component
systems.} }

\subsection{Perform Dynamical Interactions}
\label{sec:dynamics}

\begin{table*}[tb]
\centering
\begin{tabular}{ccccccc}
\hline\hline
\multicolumn{4}{c}{\textbf{Initial Conditions}} & \multicolumn{3}{c}{\textbf{Runtimes (Hours)}} \\
\hline\hline
Model Name & $M_\text{BH}/M_\text{star}$ & $m_\text{BH}/m_\text{star}$
& $T_{\text{end}}$ $(t_{dyn})$ & \texttt{CMC} & \texttt{NBODY6} &
\texttt{RAPID} \\\hline
64k-0.01-10 & 0.01 & 10 & 14000 & 5.0$\pm1.7$ & 425 & 1.8$\pm0.2$\\
64k-0.02-10 & 0.02 & 10 & 20000 & 8.4$\pm0.9$ & 304 & 2.6$\pm0.3$\\
64k-0.01-20 & 0.01 & 20 & 10000 & 2.6$\pm0.4$ & 554 & 1.3$\pm0.2$\\
64k-0.02-20 & 0.02 & 20 & 20000 & 5.3$\pm0.7$ & 512 & 2.7$\pm0.6$\\\hline\hline
\end{tabular}
\caption{The model names, initial conditions, and runtimes for each of the four idealized
GC models.  The runtimes quoted are the relevant walltimes for each
method.  {We list the single runtime for each \texttt{NBODY6} run, and
the mean and standard deviation of the 10 \texttt{CMC} and \texttt{RAPID} runs for each
cluster.}  The \texttt{CMC} and \texttt{RAPID} models were run on 4 Intel Xeon E5-2670 Sandy Bridge
processors, while the \texttt{NBODY6}
models were run on 8 processors and 1 Nvidia Tesla M2090 GPU.}
\label{tab:clusters}
\end{table*}

After selecting the global simulation timestep, the \texttt{CMC} processes apply the standard nearest-neighbor interactions (two-body relaxations, strong encounters, and collisions) to successive pairs of particles in the radially sorted array of all cluster particles.  The interactions are identical to those employed in our previous \texttt{CMC} studies.  What differs in the \texttt{RAPID} approach is how the outcomes of interactions between \texttt{CMC} stars and Kira BHs are handled.  Specifically:

\begin{itemize}

\item \textbf{Star-Star} interactions are handled in the same fashion as in a pure-\texttt{CMC} integration \citep[see][and references therein]{Pattabiraman2013}.

\item \textbf{Star-BH}
interactions are also handled in the same fashion (by \texttt{CMC}); however, in addition to updating the local BH information stored in \texttt{CMC}, the dynamical changes to
each particle are communicated back to the Kira process once the interaction step is complete.

\item \textbf{BH-BH}
nearest-neighbor interactions are skipped, since such
encounters will be performed with greater accuracy in the $N$-body integration.

\end{itemize}

\noindent In \texttt{CMC}, two-body relaxations are performed by setting
nearest-neighbor stars along randomly selected hyperbolic orbits that are consistent
with their radial and tangential velocities.  The hyperbolic encounter
modifies the velocities of both stars, allowing for an exchange of energy and
angular momentum.  At the end of the timestep, we communicate the full 3D
velocity changes for each BH to the Kira process. The velocity of each particle
in the $N$-body is updated by adding the full-3D velocity perturbation to
previous 3D velocity of the BH.

In addition to two-body relaxations, \texttt{CMC} integrates strong scattering encounters between neighboring multiples (such as binary-single neighbors
or binary-binary neighbors) using the \texttt{Fewbody} small-$N$ integrator
\citep{Fregeau2004}.  In \texttt{RAPID}, we also allow for strong encounters between neighboring stars and
BHs.  However, as Kira can
produce BH higher-order multiple systems, (triples, quadruples, etc), we have
modified \texttt{Fewbody} to perform scatterings between systems of arbitrary
multiplicity, such as single-multiple and binary-multiple encounters.  We ignore
multiple-multiple scatterings, since \texttt{CMC} does not track higher-order stellar triples, and BH-BH encounters are performed naturally
in Kira.

Once \texttt{Fewbody} has completed the scattering (which can take several
seconds of CPU time for compact higher-order multiple systems) the output is
then sent back to \texttt{CMC} and Kira separately.  For higher-order multiples, the full
3D position and velocity of each BH component, relative to the multiple center-of-mass,
is sent to the Kira process.  The multiple is then reinserted into the $N$-body integration at its previous position.

The only exception to this procedure is the formation of mixed-multiple systems
(a binary or higher-order multiple with both BH and stellar components).  We evolve any BH-Star binaries in \texttt{CMC}.  { For higher-order multiples with
both stellar and BH components, the multiple is hierarchically broken apart into smaller components.  Any star or BH-Star is evolved by \texttt{CMC}, while any BH, binary BH, or BH multiple is evolved by Kira.
 The kinetic energies of the newly-broken components are adjusted to ensure 
 conservation of energy.} {While this limits the  modeling of BH-non BH systems (such as low-mass X-ray binaries), these 
 systems predominatly form with low mass BHs at the outer region of the BH 
 subsystem, where mixing between stars and BHs is more common 
 \citep{2018ApJ...852...29K}.  Because of this, these systems are less likely to 
 participate in the strong three-body encounters that form BH binaries in the 
 central regions of the cluster.}

\subsection{Perform Stellar Evolution} \texttt{RAPID} considers realistic stellar
evolution using the Single Stellar Evolution (SSE) and Binary Stellar Evolution
(BSE) packages of \cite{Hurley2000,Hurley2002}.  This is identical
to the previous implementation in \texttt{CMC}.  No stellar evolution is required by the direct $N$-body, since the only particles integrated by Kira are BHs. As
stated above, all mixed BH-Star objects are integrated in \texttt{CMC}. This is to ensure that the binary stellar evolution for BH-star systems is performed consistently.

\subsection{Calculate New Orbits and Positions}
\label{sec:orbits}

After the dynamical information for each particle has been updated, a new
orbital position and velocity, consistent with the particle's new energy and angular momentum, must be
computed.  Since the dynamical state of each particle is up-to-date in \texttt{CMC} and Kira, the MC and $N$-body integrations can be performed in parallel by their respective processes.

\subsubsection{Orbit Calculation (MC)}
\label{sec:orbits:MC}

Orbits in \texttt{CMC} are determined using the standard H\'enon-style
orbit-sampling approach.  We assume that the orbit is entirely a function of the
star's kinetic energy and the spherical potential of the cluster.  We first
constrain the radial extent of the orbit by computing the two zeros
($r_{\text{min}}$ and $r_{\text{max}}$) of the energy equation:

		\begin{equation} 2E - 2 \Phi(r) + J^2 / r^2 = 0 \end{equation}

		\noindent Then, we probabilistically select a new radius for the star
		based on the star's new orbit.  Since the probability of finding the
		star at a given radius is proportional to the time the particle spends
		at that radius, we can express the probability of finding the star at a
		specific radius as

		\begin{equation} P(r) dr = \frac{dt}{T} =
		\frac{dr/|v_r|}{\int^{r_{\text{max}}}_{r_{\text{min}}} dr / |v_r| }
		\end{equation}
\noindent We then draw a random sample from $P(r)$, and use the value as the
particle's position for the next timestep.

\subsubsection{Orbit Calculation (N-body)}
\label{sec:orbits:nb}

Because we have dynamically perturbed the BHs though scattering and a new external potential, the gravitational force and its higher derivatives must be recalculated before resuming the Kira integration.  Otherwise, the dynamical changes to the BH velocities would produce discontinuities in the higher derivatives of the force, breaking the smoothness needed for the $4^{\text{th}}$-order Hermite integrator to work.  The $N$-body system is reset using Kira's built-in reinitialization function, which recomputes the acceleration and jerk for each BH explicitly.
The position and velocities of all the particles (up to any changes from dynamical interactions) are not modified.

Once the system has been reinitialized, the $N$-body configuration is directly
integrated with the Kira integrator for a number of dynamical times equal to the current \texttt{RAPID} timestep.  For each particle, the total force is computed as the sum of the external
force, from $\Phi^{\text{MC}}$, and the internal force
computed via direct summation of the other BHs.  We apply the external \texttt{CMC} potential to each
isolated BH and the center-of-mass of each bound multiple system by computing the
acceleration and jerk from the external potential

\begin{align*} \vec{a}_{\text{ext}} &= -\frac{\partial
\Phi^{\text{MC}}}{\partial r} \hat{r}\\ \vec{j}_{\text{ext}} &= -\frac{1}{r}
\frac{\partial \Phi^{\text{MC}}}{\partial r}(\vec{v} - (\vec{v}\cdot
\hat{r})\hat{r}) - \frac{\partial^2 \Phi^{\text{MC}}}{\partial
r^2}(\vec{r}\cdot\hat{r})\hat{r}  \end{align*}

\noindent and adding this to the acceleration and jerk of each BH using Kira's external potential module.
To simplify the calculation of the potential and its derivatives, we select 30
stars evenly in $\log r$ from the innermost \texttt{CMC} star to the stripping radius of
the $N$-body simulation, and copy their radii and $\Phi^{\text{MC}}$ values
to the $N$-body integrator.  The values of $\Phi^{\text{MC}}$ and its radial
derivatives are then calculated via $5^{th}$-order Lagrange Polynomial interpolation.
  This technique is inspired by the radial potential sampling used in \cite{Vasiliev2014a}.

\subsection{Sort Radially}
\label{sec:sort}

Finally, once all the relevant physics has been applied and the data collected on the \texttt{CMC} processes, the particles must be sorted in order of increasing radial distance from the GC center.  The sorting is performed in parallel by all \texttt{CMC} processes using the parallel Sample Sort algorithm described in \cite{Pattabiraman2013}.

\section{Parallelization Strategy}
\label{sec:parallelization}

To incorporate the Kira integrator into \texttt{CMC}, we make the
following modifications to our parallelization strategy. When the simulation starts, all
particles are evolved using the \texttt{CMC} scheme.  As described in Section
\ref{subsec:partition}, the activation criterion for the $N$-body integrator
depends on the type of particles being integrated.  For point particle
simulations, the $N$-body integrator is begun immediately, whereas for star
clusters modeled with stellar evolution, the $N$-body integrator is only
activated once a certain number ($\sim 25$) of BHs have been formed. Once the activation condition is met, we divide the
entire set of particles into two sets: MC stars and $N$-body BHs.

At the same time, we divide the
existing set of $p$ processes into two separate groups: a single Kira process
for integrating the BHs,
and $p-1$ \texttt{CMC} processes for integrating the stars.  Any MPI communication is handled by
two custom intracommunicators: one corresponding to all $p$ processes and one
restricting communication to the $p-1$ \texttt{CMC} processes.  The latter allows
us to employ the same parallelization strategy described in
\cite{Pattabiraman2013} with minimal modification.  When the processes are split,
all the BHs are sent to the Kira process, where their coordinates are
converted from $(r,v_r,v_t)$ space to the full 6-D phase space
by randomly sampling the orientation of the position and velocity vectors.  This
sets the initial conditions for the $N$-body
integration.  In addition, the Kira process maintains a radially-sorted
array of $(r,v_r,v_t)$ for all BHs.  This is done
to facilitate easier communication with \texttt{CMC} and to
minimize the required MPI communication between \texttt{CMC} and Kira
every timestep. Although the \texttt{CMC} processes hand over their BHs to the Kira process,
the BHs are not deleted from their local arrays. The
BHs are left intact yet inert, so that their positions and velocities
can be updated upon completion of the $N$-body integration.

Once the \texttt{CMC} and Kira MPI processes have been initialized,
the hybrid method must allow both codes to interact while minimizing the amount
of MPI communication.  This is accomplished by a series of intermediate arrays,
designed to store and transmit the minimum amount of information back and forth between the
\texttt{CMC} and Kira.  Communication only occurs at two points
during a \texttt{RAPID} timestep.  The first occurs after the relaxation and strong-encounters
have been performed, in order to communicate the dynamical changes from \texttt{CMC} to Kira.
The second occurs after both systems have completed their respective
orbit computations, to communicate the new dynamical positions and velocities from Kira back to \texttt{CMC}.

\subsection{\texttt{CMC} to Kira Communication}

After the \texttt{CMC} processes have computed a new potential, performed
two-body relaxations, and any strong encounters, the
dynamical changes to the BHs must be communicated to the Kira process.  This is done with three distinct communications:

\begin{itemize}
\item The \textbf{MC potential}, $\Phi^{\text{MC}}$, is sent to the
Kira process as two arrays, containing the radius and cluster potential of every
star (excluding the BHs) in \texttt{CMC}.  The
Kira process selects 30 of these stars as described in Section
\ref{sec:orbits:nb}, and passes the information to the Kira
integrator to use when computing the external force.

\item The \textbf{two-body relaxations} are communicated as an array of objects,
each containing a particle ID and a 3-dimensional $\Delta \vec{v}$.  These weak
velocity perturbations are added to the single BHs and the centers-of-mass of
any BH multiple systems before the $N$-body system is reinitialized by Kira.

\item The results of \textbf{strong encounters} are communicated differently
depending on the type of encounter.  For binary BHs that have experienced a
strong encounter with a star, only the change in semi-major axis and
eccentricity are communicated back to Kira.  For triples and
higher-order multiples, the full position and velocity of every
BH in the multiple is communicated to Kira.  To reduce communication, any
hierarchical information is not transmitted, and the Kira process reconstructs
the hierarchy locally before the $N$-body system is reinitialized.
\end{itemize}

Since all the information previously described does not drastically change the
radial positions or velocities of the particles in the $N$-body (by assumption,
the MC approach requires that $\Delta v / v \ll 1$), the dynamical
state of the $N$-body system is preserved between \texttt{RAPID} timesteps.  The one exception
is strong encounters in which a single bound BH multiple is broken into
components.   Since strong encounters in \texttt{CMC} are performed by
assuming the scatterings are isolated from the cluster potential at infinity, the resultant
components cannot be placed at the same infinite location in the $N$-body system.  For
such systems, the components are placed at the correct radius and random orientations, similar to the
initial transfer of BHs from \texttt{CMC}.

Finally, we allow for the possibility that \texttt{CMC} may create new BHs,
either through stellar evolution or through strong
encounters which produce single or binary BHs.  We add any such new BHs to Kira.  The new BH is then
flagged as a BH in the local array of the \texttt{CMC} process that created it, to ensure it
is not evolved by \texttt{CMC} during the next timestep.

\subsection{Kira to \texttt{CMC} Communication}

After the Kira integrator has computed the orbits of the
BHs, the new dynamical state must be communicated back to the \texttt{CMC} processes.  First, the dynamical 6-D phase space information for each
particle in the $N$-body is projected back to the reduced $(r,v_r,v_t)$ basis,
and copied back into place in the intermediate star array on the Kira
process.  The intermediate BH array is then divided up and communicated back to the respective \texttt{CMC}
processes.

In addition to the positions and velocities, any new objects, such as single
BHs, newly formed binaries, or higher-order
multiples, are communicated as new objects to the last ($p-1$) \texttt{CMC}
processes, to be placed in the correct process once the \texttt{CMC} (and
intermediate BH) arrays are sorted.
For single and binary BHs, this is accomplished by sending the usual
$(r,v_r,v_t)$ for each system and the semi-major axis and eccentricity for any
binaries to the \texttt{CMC} processes.  For higher-order multiples, the full 6D
dynamical information is transmitted back using the same array that sent the
strong encounters from \texttt{CMC} in the first communication.  Again, only the
positions and velocities of the BH multiple components are communicated, so the hierarchical information is reconstructed on the local \texttt{CMC} process after communication.

Because the $N$-body integration is being performed in a lower-density environment than the full cluster, it is possible for Kira to produce pathologically wide binaries and higher-order multiples.  This effect is particularly problematic at late times, where a handful ($\sim 10$) of BHs can easily produce binaries with separations greater than the local inter-particle separation of stars.  Since the unphysically large interaction cross-sections of these systems drastically shrink the \texttt{CMC} timestep, we break apart any multiple systems whose apocenter distance is greater than 10\% of the local inter-particle separation of stars at that radius.  The kinetic energy of the \texttt{CMC} stars is adjusted to ensure energy conservation.  This criterion for breaking wide binaries is the same criterion used to break wide binaries produced by \texttt{Fewbody} in \texttt{CMC}.

\section{Analytic Comparison}
\label{sec:plummer}

\begin{figure}[]
\centering
\includegraphics[scale=0.65]{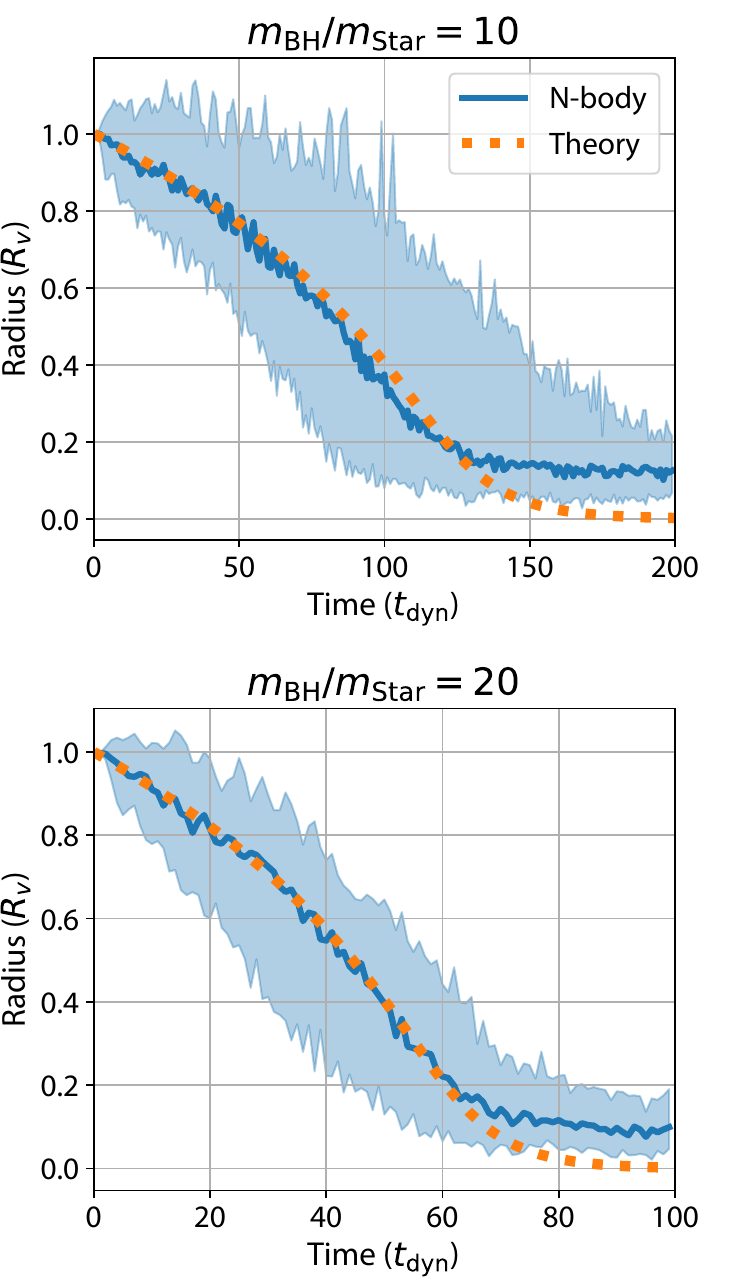}
\caption{\textbf{Dynamical Friction on a Single Particle ($N$-body)} -- The inspiral of a single massive BH in a Plummer sphere due to
dynamical friction, as computed by integrating equation \ref{eqn:drdt} (in dashed-orange) and by direct
$N$-body (in blue, using the Kira integrator).  The $N$-body results are
averaged over 50 independent realizations, with the solid blue line and the blue
shaded region indicating the median and 90-percentile values of the radius
when binned every 1 dynamical time.  We consider BHs with masses 10 and 20 times
the mass of the stars (top and bottom, respectively).}
\label{fig:dynFricNbody}
\end{figure}

\begin{figure}[]
\centering
\includegraphics[scale=0.65]{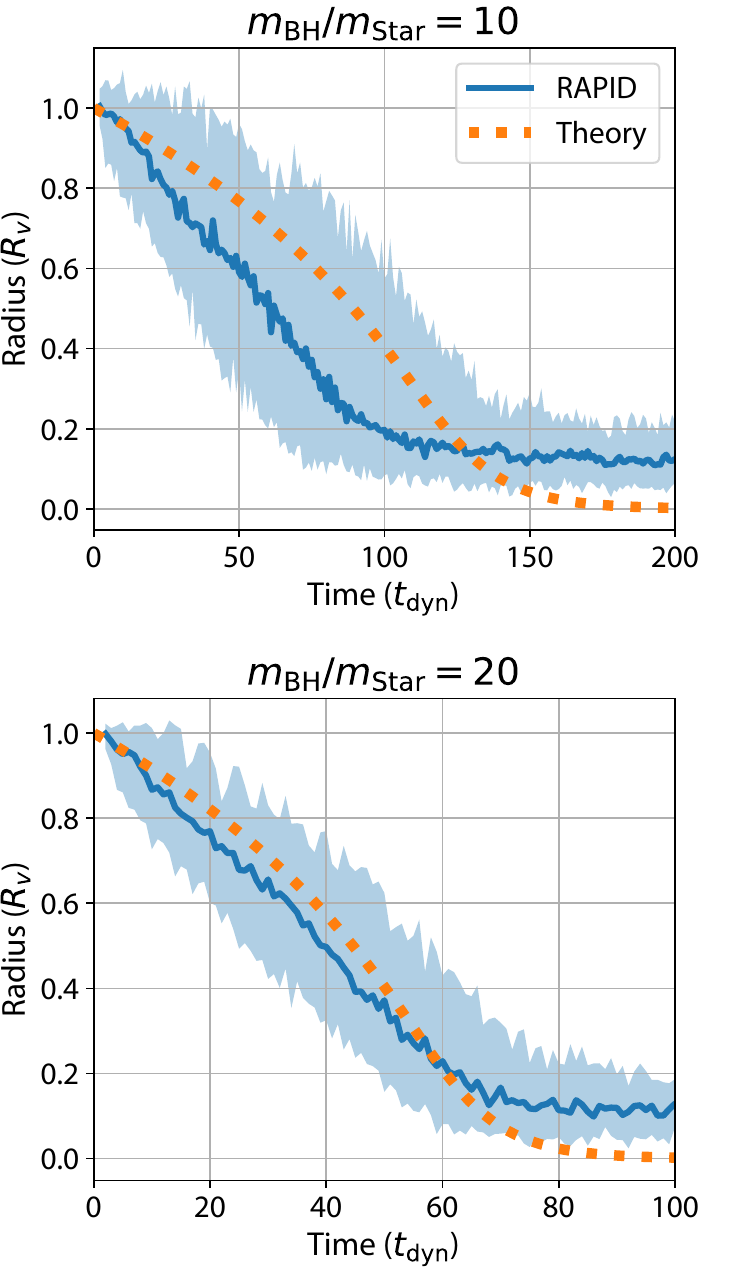}
\caption{\textbf{Dynamical Friction on a Single Particle (\texttt{RAPID})} -- Similar to Figure 4, but with the numerical
results from \texttt{RAPID}.  Here the agreement between theory and numerical
experiment is somewhat diminished, particularly for the
$m_{\rm{BH}}/m_{\rm{star}} = 10$ case, where the different in inspiral times can
be as high as 30\%; however, the agreement improves considerably for more
massive BHs.  }
\label{fig:dynFricRapid}
\end{figure}

To test our new method, we first choose a simplified analytic model designed to
demonstrate whether \texttt{RAPID} can model the dynamical friction experienced by a
single BH in a sea of stars.  Theoretically, the time taken for a massive object to spiral in to the
center of a cluster can be calculated considering the rate of change of angular
momentum of the particle due to the dynamical
friction experienced by said particle
\citep{Chandrasekhar1943,Binney2008}.  Consider a particle of mass $m$ on a
circular orbit at a radius $r$
in a Plummer sphere of mass $M$ and scale factor $a$ \citep[see
e.g.,][]{Heggie2003}.  Many properties of the Plummer sphere can be described
analytically, such as the mass interior to $r$

\begin{equation}
M(r) = M\left( 1+\frac{a^2}{r^2} \right)^{-3/2}
\end{equation}

\noindent the density at a radius $r$

\begin{equation}
\rho(r) = \frac{3M}{4\pi a^3}
\left( 1+\frac{r^2}{a^2} \right)^{-2.5}
\label{eqn:plumRho}
\end{equation}

\noindent and the velocity dispersion at $r$

\begin{equation}
\sigma(r) = \sqrt{\left( \frac{G M}{a} \right)\sqrt{1+\frac{r^2}{a^2}}}
\end{equation}

The specific angular momentum of our massive particle is given by $L =
rV_{c}$, where $V_c = \sqrt{G M(r)/r}$ is the circular velocity at radius $r$.  The rate of change of angular momentum is then

\begin{align}
\frac{dL}{dt} &= \left( V_c + r \frac{dV_c}{dr} \right)\frac{dr}{dt}\nonumber\\
&= \frac{V_c}{2} \left(1+ 3\left( 1+\frac{r^2}{a^2} \right)^{-1} \right)
\frac{dr}{dt} \label{eqn:angMom}
\end{align}

\noindent As the BH travels through the cluster, it experiences dynamical
friction in the opposite direction of its velocity.  If we assume that the
change in $r$ is sufficiently slow that the BH orbit remains circular,
then the change in angular momentum due to dynamical friction is given by

\begin{align}
\frac{dL}{dt}\Bigr|_{\rm{fric}} &= a_{\rm{fric}} r\nonumber\\
&= -4 \pi G^2 \log\Lambda \rho(r) \chi  m r V_c^{-2} \label{eqn:dynFric}
\end{align}

\noindent where $_{\rm{fric}}$ is the acceleration due to dynamical friction \cite[][p.~645]{Binney2008}  $\chi \equiv  \erf(X) - 2 X \exp(-X^2)/\sqrt{\pi}$, $X \equiv V_c / (\sqrt{2} \sigma(r))$, and $\log\Lambda \sim \log\gamma N$ is the Coulomb Logarithm, with $N$ being the number of particles in the cluster and $\gamma\sim 0.01$.  Setting equations \ref{eqn:angMom}
and \ref{eqn:dynFric} equal yields

\begin{equation}
\frac{dr}{dt} = \frac{- 8 \pi G^2 \log\Lambda \rho(r) \chi  m r }{V_c^3  \left(1+ 3\left( 1+\frac{r^2}{a^2} \right)^{-1} \right) }
\label{eqn:drdt}
\end{equation}

\noindent which can be solved numerically for $r(t)$.

To test whether our proposed method can reproduce the analytic inspiral times
predicted by equation \ref{eqn:drdt}, we create 50 independent realizations of a
Plummer sphere by drawing $10^4$ equal-mass particles from equation
\ref{eqn:plumRho}.  We then place a single massive particle on a
circular orbit at the virial radius of the cluster.  We consider mass ratios
between the BH and the individual stars of 10 and 20.  Each cluster model is then
integrated forward until the particle has settled into the center.  We
integrate these models forward using the Kira integrator (Figure
3) and the new \texttt{RAPID} approach (Figure
4).

Because \texttt{RAPID} and \texttt{CMC} are designed to model two-body relaxation by
averaging various quantities over several neighboring stars, special care must
be taken for \texttt{RAPID} to accurately model the behavior of a single particle.  To
that end, we set the maximum scattering angle to the typical value of
$\theta_{\rm{max}} = \pi / 2$,  while we reduce the number of neighboring
particles over which the quantities in equation \ref{eqn:trel} are averaged to 2.  In
other words, the quantities used to compute the timestep consider only the
nearest particles when computing the local two-body relaxation timescale
(while the global timestep is chosen as the minimum of all nearest neighbor
timescales).  This ensures
that the two-body relaxation timestep chosen for the \texttt{RAPID} simulations can
appropriately model the dynamical friction experienced by a single massive
object.

As expected, the pure $N$-body agrees with the theoretical prediction to a
better degree than \texttt{RAPID}.  However, given the
approximate nature of two-body relaxation in \texttt{RAPID}, the agreement presented in
Figure 5 is encouraging, and suggests that our
treatment of two-body relaxation between Monte Carlo and $N$-body can treat
dynamical friction to within $\sim 30\%$ (the largest deviation between theory
and numerical results in the $m_{\rm{BH}}/m_{\rm{star}} = 10$ case), with
substantial improvements for more massive BHs (the $m_{\rm{BH}}/m_{\rm{star}} =
20$ case being more representative of true BH masses in realistic clusters).

We reiterate that these results were obtained by reducing
the number of particles used to compute various average quantities (Section
\ref{subsec:timestep}).  This allows the
timestep to be properly calibrated for the dynamical friction of a single
particle (the massive BH).  In a standard \texttt{CMC} and \texttt{RAPID} run, a larger
averaging kernel can be used, since for realistic clusters with a continuous
mass function, there will be many massive and light objects within the inner-most 40
particles (which typically sets the minimum relaxation time for the cluster).  For more idealized clusters,
where the mass function is discrete and there can sometimes be only one massive
BH per 40-particle kernel (especially before mass segregation), the averages must be computed carefully.
In this section, we have accomplished this by setting $\theta_{\rm{max}}$ to the standard value of
$\pi / 2$ and taking the mimimum timestep computed over averages between the nearest-neighbor particles
(usually the average between the BH and its two neighboring stars).  More
generally, this reduction in timestep can also be accomplished by reducing the
maximum angle for a two-body deflection to
$\theta_{\rm{max}}=1$.  We elect for the later in the next section, as it has
been demonstrated to work well for the idealized two-component
systems considered there \citep{Fregeau2007}.

As an aside, it is interesting
to note that Figure 4 provides an excellent way to test the
value of $\gamma$ commonly used in numerical work involving the Coulomb
Logarithm ($\log\Lambda \equiv \log{\gamma N}$).  Even small changes (such as
$\gamma = 0.005$ or $\gamma=0.02$) produce obviouly worse agreement in Figure
3.  This suggests that the value of $\gamma=0.01$ used in
 many previous studies of multi-mass clusters is appropriate.

\begin{figure*}[tb!]
\caption{\textbf{Half-mass and Core Radii of Idealized Clusters} -- The half-mass and core radii, {averaged over 100 dynamical times}, for all four models as determined by
direct $N$-body (\texttt{NBODY6}, in black), the Monte Carlo (\texttt{CMC}, in
blue), and the hybrid approach (\texttt{RAPID}, in red).  {We only
show one realization from \texttt{NBODY6}, while we show 10 realizations each
for \texttt{CMC} and \texttt{RAPID}}  {\texttt{RAPID} reproduces both the core and
half-mass radii at early times better that \texttt{CMC} (with the
exception of the 64k-0.01-10 run, though \texttt{RAPID} is still much 
closer than \texttt{CMC}).  On the other hand, \texttt{RAPID} reproduces the core radii of
the full $N$-body runs far more accurately than \texttt{CMC}, with the core radii from
\texttt{NBODY6} and \texttt{CMC} disagreeing by up to a factor of 3.}
 At late times, the core
radii from \texttt{RAPID} begins to diverge from \texttt{NBODY6}; this
occurs when the majority of the BH system
has been ejected, and the \texttt{CMC}-controlled two-body relaxation dominates the BH dynamics.  \texttt{RAPID}
reverts to pure \texttt{CMC} when the number of BHs drops below a certain threshold (5
BHs, indicated by the black dashed lines, (with the gray bands indicating
the $1\sigma$ variation across the 10 runs).}
\centering
\includegraphics[width=0.95\textwidth]{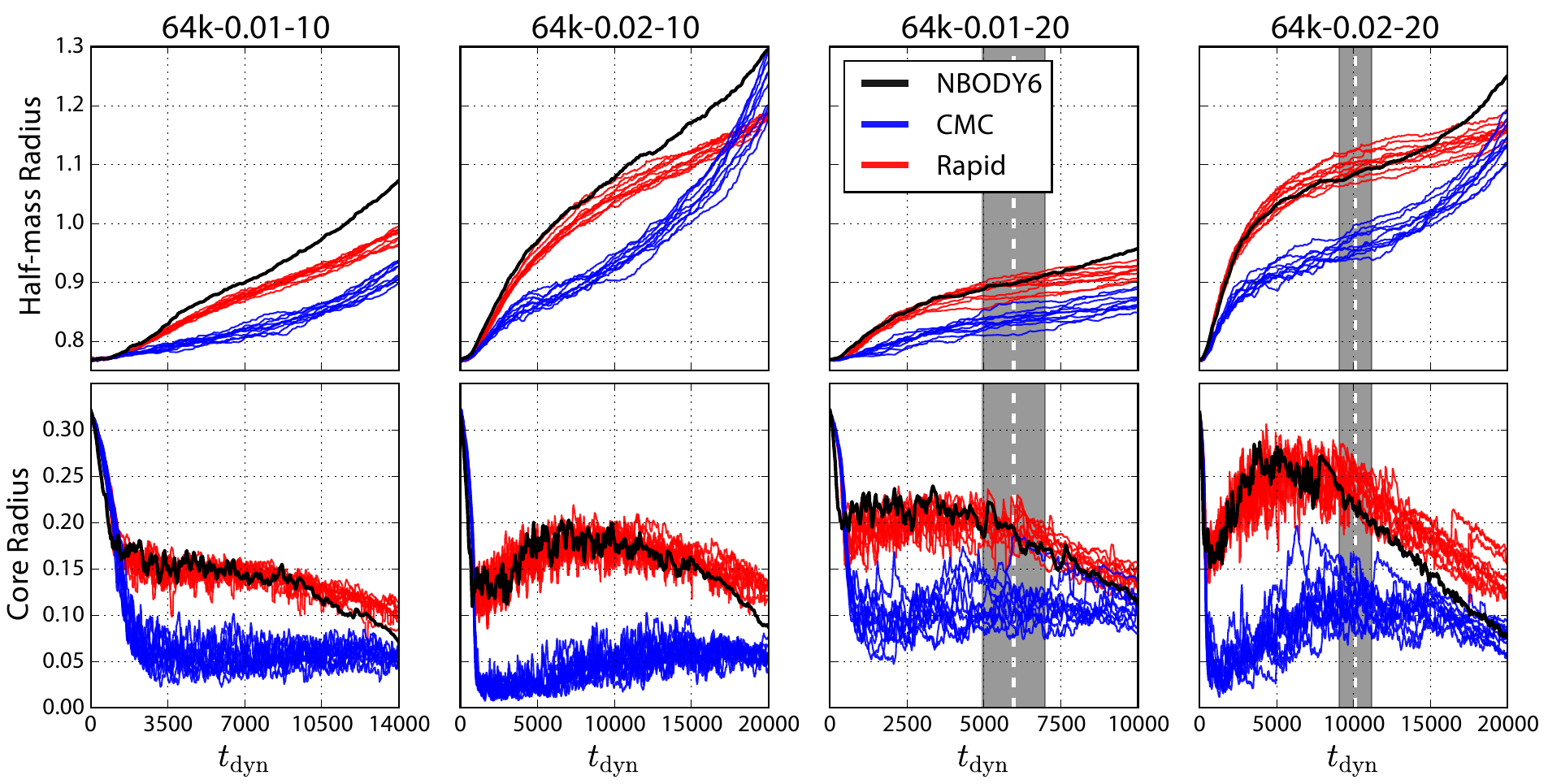}
\label{fig:rcrh}`
\end{figure*}

\begin{figure*}[tb!]
\caption{\textbf{Phase-space Scatterplots of Stars and BHs} -- The specific energy (potential plus kinetic) versus the specific angular
momenta for each particle at four different points during the evolution of the
$m_\text{BH}/m_\text{star} = 20$ and $M_\text{BH}/M_\text{star}=20$ model.  We
show the scatter plot of $E$ vs $J$ at the beginning of the simulation, at the
point of deepest collapse ($500~T_{\rm{dyn}}$), at the point of greatest core
re-expansion ($5,000~T_{\rm{dyn}}$), and at the end of the simulation ($
20,000~T_{\rm{dyn}}$) after the majority of BHs have been ejected.}
\centering
\includegraphics[width=0.95\textwidth]{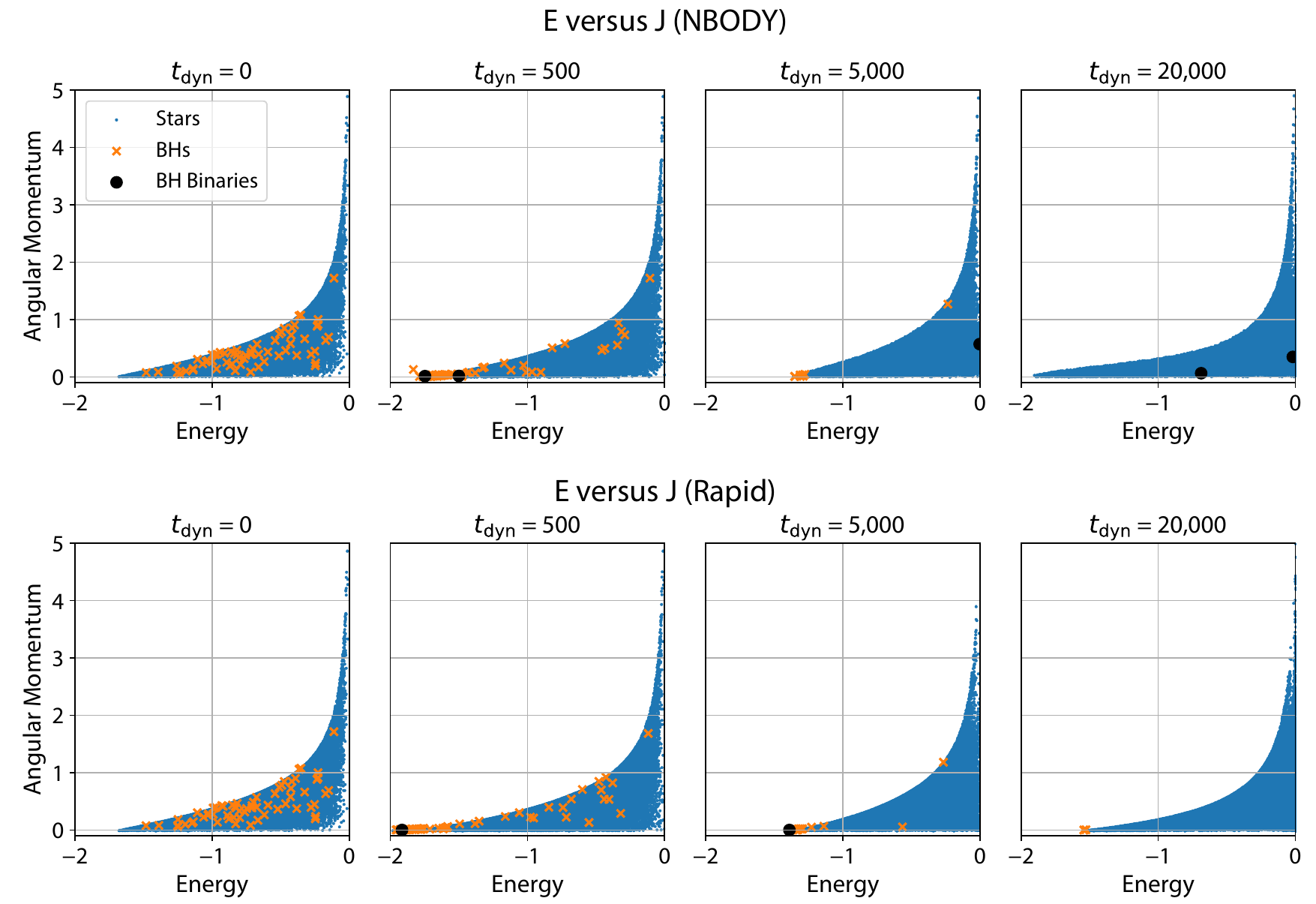}
\label{fig:evj}`
\end{figure*}
\begin{figure*}[]
\centering
\includegraphics[scale=0.65]{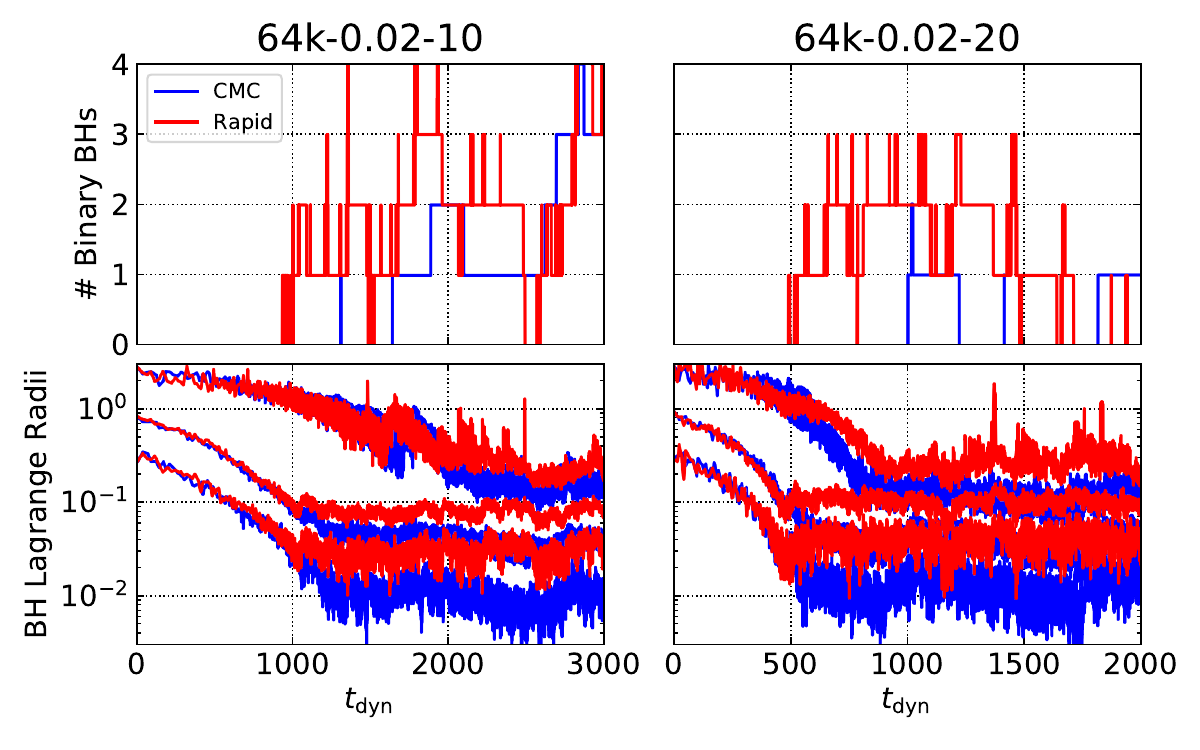}
\caption{\textbf{BH Lagrange Radii and Binary BH Formation} -- The number of binary 
BHs and the Lagrange radii for all BHs during the early stages of cluster 
collapse for two example clusters.  On the top, we show the number of BH 
binaries as a function of time in both \texttt{RAPID} and \texttt{CMC}.  On the 
bottom, we show the radii enclosing 10\% 50\% and 90\% of the BHs in the cluster 
for the two techniques.  In both cases, as the cluster collapses, \texttt{RAPID} 
forms a binary through three-body encounters much earlier, halting the continued 
collapse of the cluster.  \texttt{CMC} continues to collapse until its first binaries are formed several hundred dynamical times later.
}
\label{fig:dynFricRapid}
\end{figure*}

\begin{figure*}[t!]
\caption{\textbf{Number of Single and Binary BHs} -- {The number of BHs retained in each cluster model over time.  On the
top, we show the total number of BHs present in each of the four clusters (including those in bound multiple
systems) for the single \texttt{NBODY6} model (in black), the 10 \texttt{CMC} models
(in blue), and the 10 \texttt{RAPID}  models (in red).  In each case, the
\texttt{RAPID} show better agreement with the \texttt{NBODY6}
model than the pure \texttt{CMC} models.  In the second and third rows, we show the
total number of BH binaries and BH triples (which cannot be produced in
\texttt{CMC}) present in the \texttt{NBODY6} model
(in black) and a single \texttt{RAPID} model (in red) as a function
of time.  Although highly stochastic, the number of bound multiple systems shows
good
qualitative agreement between the two methods.  As in Figure 6, we
show the mean and $1\sigma$ times when \texttt{RAPID} reverts to a pure MC approach
with the white-dashed line and gray bands, respectively.  For the middle and
bottom rows, the white line indicated the reversion time for that run.}}
\centering
\includegraphics[width=0.95\textwidth]{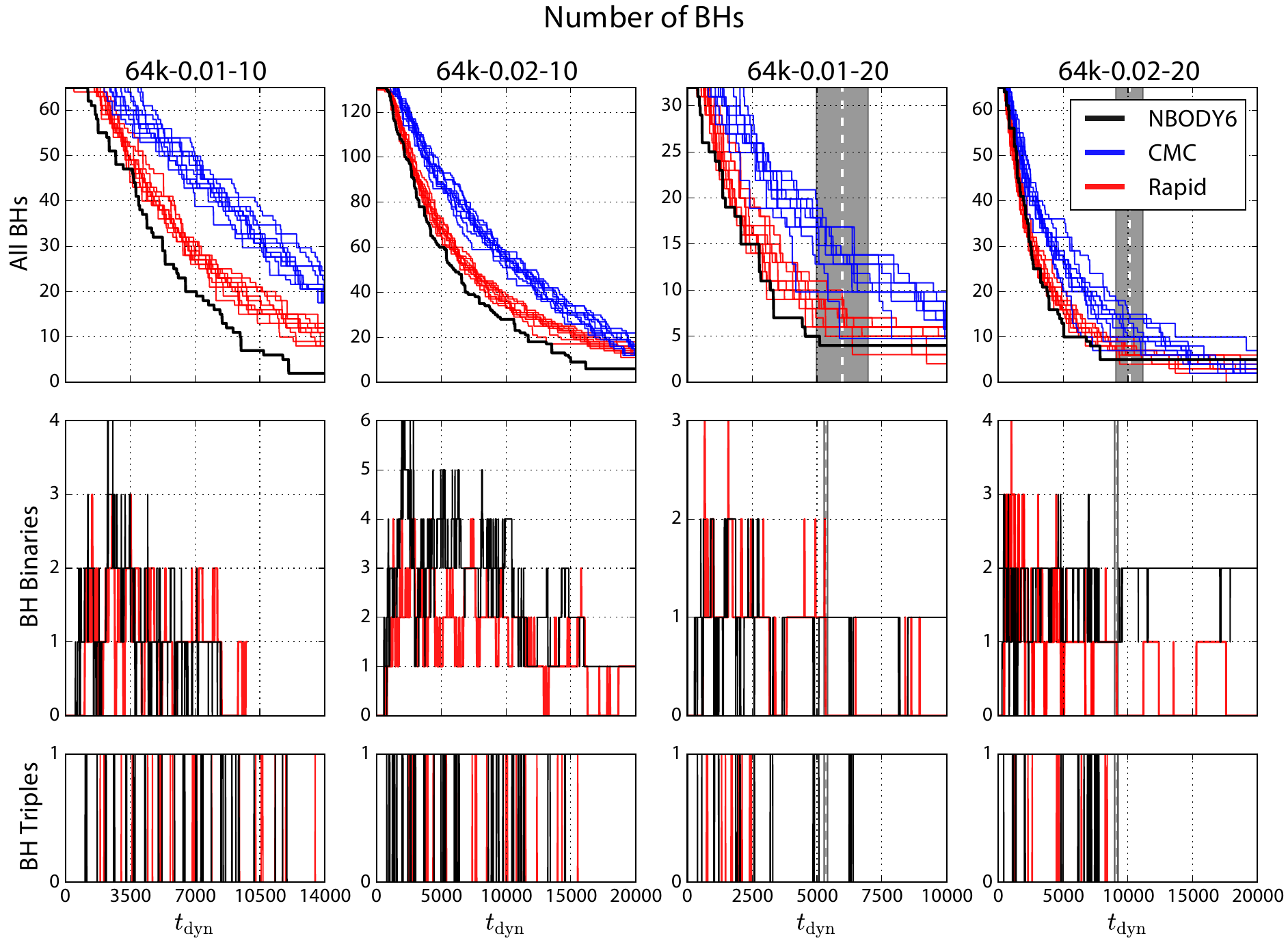}
\label{fig:nbh}
\end{figure*}

\begin{figure*}[t!]
\caption{\textbf{Properties of Ejected Binary BH systems} -- {The properties of the ejected binary BH systems from each of the four
\texttt{NBODY6} models (black diamonds) and from the 10 \texttt{RAPID} realizations
(red circles).
We show the eccentricity and semi-major axis (in AU, assuming each cluster to
have an initial virial radius of 1 parsec) for each binary.  In each case, the \texttt{NBODY6}
results align with the larger sample size from the \texttt{RAPID} models.  The text in
each plot indicates the total number of ejected binaries from the
\texttt{NBODY6} models, and the mean and standard deviation from
the 10 \texttt{RAPID} models.}}
\centering
\includegraphics[width=0.95\textwidth]{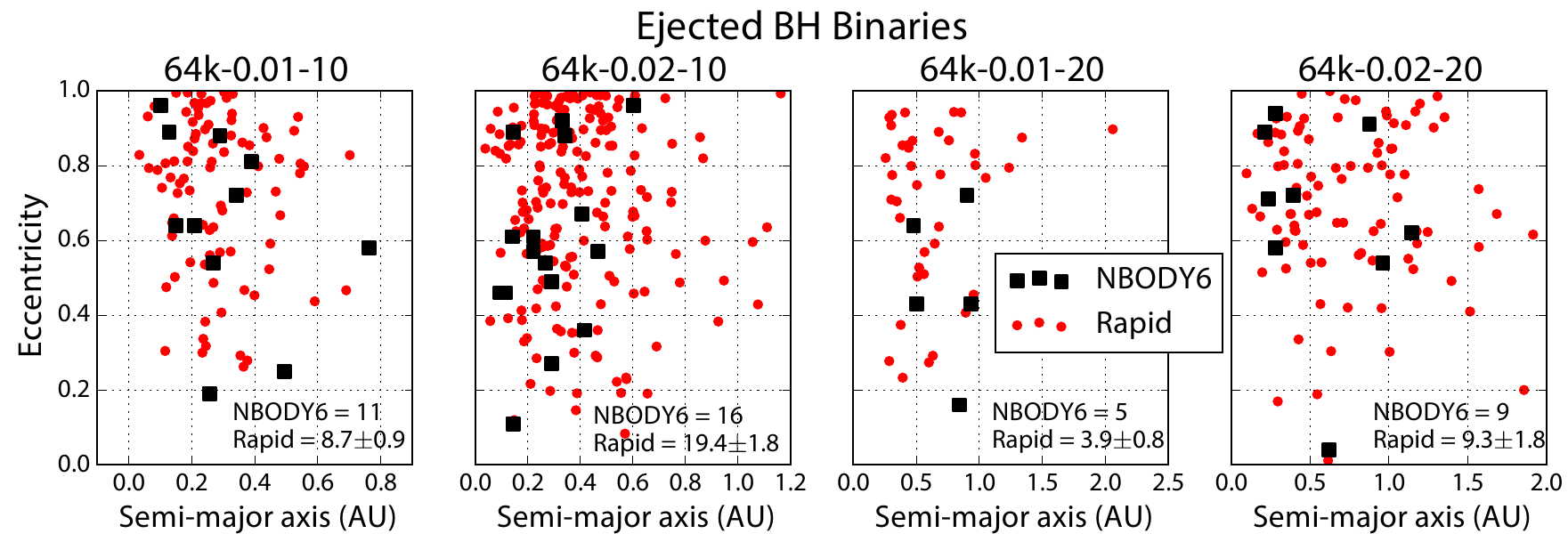}
\label{fig:bbh}
\end{figure*}

\section{Numerical Comparison}
\label{sec:results}

To test the effectiveness of this hybrid approach, we compare our code to
idealized models of GCs using similar initial conditions to \cite{Breen2013}.
We considered four clusters with 65,536 point-mass particles.  The majority of particles are low-mass stars, while a small fraction of particles are high-mass BHs.  We considered different mass ratios
between BHs and stars ($m_\text{BH}/m_\text{star} = 10$
and $m_\text{BH}/m_\text{star}=20$).  We also varied the total mass in stars and
BHs ($M_\text{BH}/M_\text{star} = 0.01$ and $M_\text{BH}/M_\text{star}=0.02$).
For each set of initial conditions, we compare 10 independent realizations of
each model produced by \texttt{RAPID} to 10 models produced by \texttt{CMC} and
by a single $N$-body model using \texttt{NBODY6} \citep{Aarseth1999}.
Each model was integrated until the majority of the BHs had been ejected from
the cluster.  A summary of the initial conditions and the average runtimes for
each approach is listed in Table \ref{tab:clusters}.

{Immeditally obvious from Table \ref{tab:clusters} is that the runtimes for 
\texttt{RAPID} are much shorter than the typical runtimes for \texttt{NBODY6}, sometimes by factors of a few hundred.  Slightly more surprising is that the \texttt{RAPID} runtimes also tend to be shorter than the runtimes for \texttt{CMC}.  This largely arises from the improved treatment of the cluster center in \texttt{RAPID}.  As will be shown in the next section, \texttt{RAPID} reproduces the less-dense core radii of the full $N$-body models better than \texttt{CMC}, with the latter producing core radii 2-4 times smaller and more compact than \texttt{NBODY6}.  Because the timestep of the MC is dominated by the relaxation in the densest region of the cluster, the more compact \texttt{CMC} models require many more timesteps to resolve the deep core collapses, resulting the longer runtimes.}

\subsection{Core and half-mass radii}
\label{sec:results:rcrh}

For the purposes of this analysis, we assume that the models produced by
\texttt{NBODY6} are the ``true'' clusters, since a direct integration
technique requires the fewest simplifying assumptions.  We wish to know how well
the models from our new hybrid technique match the models produced by the
$N$-body approach.  To that end, we focus on two typical figures of merit in star cluster simulations: the half-mass and core radii.  In both cases, the radii are defined as the distance from the center of the cluster.  For \texttt{NBODY6}, this is calculated as a density-weighted sum:
\begin{equation}
    \vec{r}_d = \left. \sum^N _{j=1} \rho_j \vec{r}_j \middle/ \sum^N _{j=1} \rho_j~, \right.
    \label{eqn:center}
\end{equation}

\noindent {where $N$ is the total number of particles, $\vec{r}_d$ is the position of each particle, and $\rho_j$ is the density of objects surrounding that particle \cite[c.f. Equation 15.1,][]{Aarseth2003}.  In \texttt{CMC} and \texttt{RAPID}, the center is fixed at $r=0$ by assumption.  The half-mass radius is the radius from the cluster center that encloses half the cluster mass.  For the core radius, all three approaches use the approximate core radius definition from \cite{Aarseth2003}, based on the unbiased estimator developed in \cite{Casertano1985}.   This takes the form of a weighted sum over distances from the cluster center out to the half-mass radius\footnote{In \texttt{NBODY6}, the sum is performed out whichever is greater of the half-mass radius or three times the previous core radius.}:}

\begin{equation}
    r_c = \sqrt{\left. \sum^{N/2}_{j=1} \rho_j^2|\vec{r}_j - \vec{r}_d|^2 \middle/ \sum^{N/2} _{j=1} \rho_j^2 \right.}~.
    \label{eqn:corerad}
\end{equation}

\noindent {In \texttt{NBODY6}, Equations \eqref{eqn:center} and \eqref{eqn:corerad} are calculated using the full 3D position vectors (with the densities for each particle computed with respect to its three nearest neighbours), while for \texttt{CMC} and \texttt{RAPID}, Equation \eqref{eqn:corerad} is calculated from the 1D positions where the densities are calculated for each particle by averaging over its 40 nearest neighbors.}

In Figure 6, we compare the half-mass radius
and core radius for each model as determined by \texttt{CMC}, \texttt{NBODY6},
and \texttt{RAPID}.  While the half-mass radii are relatively consistent
across all models {(with a maximum deviation of $\sim 10\%$)}, the core radii are drastically different between the three
methods.  \texttt{CMC} consistently underestimates the core
radius of each cluster immediately after core collapse, with the trend
persisting until all BHs have been ejected from the cluster.  This consistent
underestimation of the cluster core radius has been observed in other studies
using \texttt{CMC} \citep{Morscher2015} and other orbit-sampling Monte Carlo
codes.  However, the core radii as determined by \texttt{RAPID} agree very
well with the radii reported by the direct $N$-body.  This suggests
that the \texttt{RAPID} approach can correctly model the
dynamics of the massive BHs which dominate the long-term evolution of GC cores.

At late times, both the half-mass and core radii
predicted by \texttt{RAPID} begin to diverge from those determined by \texttt{NBODY6}.
This divergence is to be expected: as the BHs are ejected from the
cluster, the orbits of individual BHs are determined less by their encounters
with other BHs, but by two-body relaxation with stars controlled by the MC.
Given the disagreement between the pure \texttt{CMC} and \texttt{NBODY6}, this
divergence is consistent. This suggests that \texttt{RAPID} will be most effective
when modeling systems that retain a large number of BHs. Recent work 
\citep[e.g.,][]{Mackey2007,Downing2012,Morscher2012,Morscher2015,Kremer2018,Askar2018} has shown
that the most massive GCs can retain hundreds to thousands of BHs up to the
present day.  Given that, \texttt{RAPID} should be able to correctly model realistic GCs
throughout their entire evolution far more accurately than a
traditional MC method.

{
In addition to the bulk evolution of the system, we also want to compare the
behaivor of individual stars in energy-angular momentum space.  In Figure 7, we plot
the specific energy and angular momentum of every star and BH in a single model for
the $m_{\rm{BH}}/m_{\rm{star}}=20$, $M_{\rm{BH}}/M_{\rm{star}}=20$ cluster.
Both models start with identical initial condition, and we show the evolution of
$E$ and $J$ as a function of cluster time.  After 500 dynamical times,
the point of deepest core collapse, the \texttt{RAPID} model lags behind the
\texttt{NBODY6} model, having produced only one BH binary, while the two BH binaries in
the \texttt{NBODY6} cluster have started to push the stars out of the central region
towards higher $E$ and $J$.  This is consistent with the slightly faster
collapse and evolution of the core radius described in Figure 6.
After 5,000 dynamical times, both models have ejected a number of
BHs, creating sufficient energy to push stars out of lowest potential energy
states in the central region.  The final snapshot (at 20,000 dynamical times)
shows the stars of the \texttt{NBODY6} model in a state of deeper collapse than
the \texttt{RAPID} model.  This is most likely due to a recent encounter between the
two remaining BH binaries pushing both onto a higher orbit in the cluster
with a correspondingly larger $E$ and $J$.  The \texttt{RAPID} model retains 3 BHs at the final snapshot.  As these remain in the cluster core,
they manage to exclude the stars from occupying the lowest energy states in the
cluster potential.  This difference at late times is consistent with the
stochastic nature of BH retention observed in more realistic cluster models.
}

\subsection{Binary Formation}
{What explains the substantial improvement in the \texttt{RAPID} core radii? In 
\cite{Rodriguez2016a}, we explored a direct comparison between \texttt{CMC} and 
a state-of-the-art direct $N$-body simulation of $10^6$ particles \citep[the 
DRAGON simulation,][]{Wang2016}.  There, we found good agreement between most of 
the structural parameters of the two cluster models (e.g., the half-mass radii, the formation and ejection rate of BHs, etc.).  However, the one notable exception was the evolution of the inner parts of the cluster such as the core radii and the inner-most Lagrange radii (the radius enclosing a certain fraction of the cluster mass).  This was especially true when considering the Lagrange radii of only the BHs \citep[][Figure 7]{Rodriguez2016a}, where the inner-most few BHs would fall into a much deeper state of collapse (by nearly two orders of magnitude) into the cluster center than the equivalent radii from the $N$-body model.  The cluster would remain in this deep state until the formation of a BH binary, which would reverse this deep collapse and bring the inner Lagrange radii back into agreement with the $N$-body results.}

{It was speculated that the reason for this discrepancy lay in the analytic prescription that \texttt{CMC} employs to model the dynamical formation of binaries during three-body encounters of single BHs.  This prescription, from \cite{Morscher2012}, may underestimate the formation rate of BH binaries, especially given that the probability of binary formation scales as $v^{-9}$ in the local velocity dispersion.  Because these interactions typically involve only a handful of objects in the cluster center, where the standard MC assumptions of spherical symmetry and $T_{\rm rel} \gg T_{\rm dyn}$ break down, it is not obvious that \texttt{CMC}'s statistical approach to binary formation based on locally-averaged quantities can correctly model this process\footnote{{Of course, H\'enon's principle ensures that the rate of binary formation and hardening will automatically adjust to satisfy the energy flux of the cluster across the half-mass radius \citep[e.g.,][]{Breen2013}.  Because \texttt{CMC} can model the global properties of realistic clusters correctly, the rate of binary formation must be correct on a relaxation timescale, regardless of the specific implementation of three-body binary formation.  What we are interested in here is the behaivor of the inner parts of the core on a dynamical timescale, where the assumptions of the MC approach explicitly break down.}}.  This difficulty was one of the primary motivators for the development of \texttt{RAPID}: by directly integrating the BH dynamics every timestep, we can explicitly model the complicated three-body encounters between single BHs on a dynamical timescale. }

{In Figure 8, we show the early stages of collapse for two typical clusters as modeled by \texttt{CMC} and \texttt{RAPID}.  The Lagrange radii, indicating the radii enclosing 10\%, 50\%, and 90\% of the BHs, are nearly identical between the two methods during the early stages of collapse (as would be expected, since both methods model dynamical friction through two-body MC relaxation).  However, in both cases, the \texttt{RAPID} models dynamically forms binaries at much earlier stages of collapse, causing the inner-most BH Lagrange radii to rebound.  The \texttt{CMC} models, on the other hand, reach a much deeper state of collapse before forming their first binaries.}

{We believe this discrepancy is responsible for the deep collapses observed in \cite{Rodriguez2016a}.  In those models, the most massive objects would naturally find themselves in the center, and continue to collapse until a binary was formed.  This caused the deep collapses noted there, which were not reproduced in the direct $N$-body model.  Here, the deep collapses have smoothed out to a more continuous underprediction, since the two-component models studied here have equal masses for all BHs, whereas in \cite{Morscher2015,Rodriguez2016a}, it was consistently the most massive BHs decoupling from the rest of the core that were responsible for the deep collapses. }

\subsection{BH Retention}

{Since the heating of the cluster is primarily driven by ejection of BHs
and binary hardening in the core \citep{Breen2013}, it is important to compare the retention and
binary production between the different methods.  In Figure 9 we
show the number of BHs retained in each cluster as a function of time.  We
show the total number of BHs for each of the 10
\texttt{CMC} and \texttt{RAPID} models, and the number of BH binaries and BH triples
present over time in the \texttt{NBODY6} model and a representative
\texttt{RAPID} model.  }

{In each of the four clusters, the \texttt{RAPID} models eject BHs at a much
faster rate than the \texttt{CMC} models, in better agreement with the
\texttt{NBODY6} model.  Although the BH ejection rate for \texttt{NBODY6} is
slightly faster than the other two methods (particularly for the
$m_{\rm{BH}}/m_{\rm{star}}=10$ cases), in each case \texttt{RAPID} performs
far better than the pure MC approach.  This is consistent with the
difference in half-mass radii between \texttt{NBODY6} and \texttt{RAPID} noted in the
previous section, where systems with less-massive BHs expand faster
in \texttt{NBODY6} than in \texttt{RAPID}.  Since the overall expansion of the cluster
regulates the ejection rate of BH binaries, models that
expand more rapidly will eject BHs more rapidly.   For BHs in the cluster, \texttt{NBODY6} and
\texttt{RAPID} produce a similar number of BH binaries and BH triples over time.
This is a noticeable improvement over \texttt{CMC}, especially when
considering BH triples, which are explicitly removed from the MC integration.}

\subsection{Ejected BH Systems}


{In addition to the retained BH binaries, it is
important to compare the properties of the ejected binaries from each cluster
model.  Since the
semi-major axis of an ejected binary is inversely proportional to its
binding energy, the orbital properties at ejection are an excellent
proxy for the hardening rate of the binaries within the core.  If two cluster
models eject a similar number of binaries with comparable binding energies, the
energy production rate in the core must also be comparable between the two
models.  In
Figure 10 we show the semi-major axes and eccentricities for the four
\texttt{NBODY6} models and the 10 \texttt{RAPID} realizations of each cluster.  In
each case, the properties of the binaries ejected by the
\texttt{NBODY6} models show good agreement with those formed by the 10
\texttt{RAPID} models.}

Although we do not show it here, the distribution of orbital eccentricities of the ejected binaries strongly follows the theoretically-predicted distribution thermal distribution \citep[$p(e)\propto e$,][]{Jeans1919}
regardless of binary mass and cluster properties.  Because the gravitational-wave merger
time for binary BHs is determined by the semi-major axis and eccentricity at
ejection \citep{Peters1964},
\texttt{RAPID} will be able to model the merger rate of binary BHs from dense
stellar clusters with similar accuracy to a direct $N$-body approach.

\subsection{Energy Conservation}
\label{sec:results:energy}

\begin{figure*}[htb]
\caption{\textbf{Energy Conservation in the \texttt{RAPID} Code} -- The energy budget and overall energy conservation for the four clusters
modeled by \texttt{RAPID} shown in Fig.\ 6.  The top plot shows the
total sum of the kinetic energy (red), potential energy (green), and multiple binding
energy (purple) for all particles over time for a single representative model
of each cluster.  The energy carried
away by the kinetic energy (blue-dotted) or binding energy (blue-dashed) of escaping
particles is also shown. {In the middle plot, we show the virial ratio between the kinetic and
potential energies.} The sum of all energies is plotted in H\'enon units
(black) in the top plot,{ while the lower plot shows the fractional change in
total energy for all 10 realizations of each model.}  The
total energy conservation error is dominated by the evolution of the $N$-body
particles in a strong external potential (the slow upward trend) and occasional
strong-encounters and tightly-bound multiple systems (discontinuous jumps).
}
\centering
\includegraphics[width=0.95\textwidth]{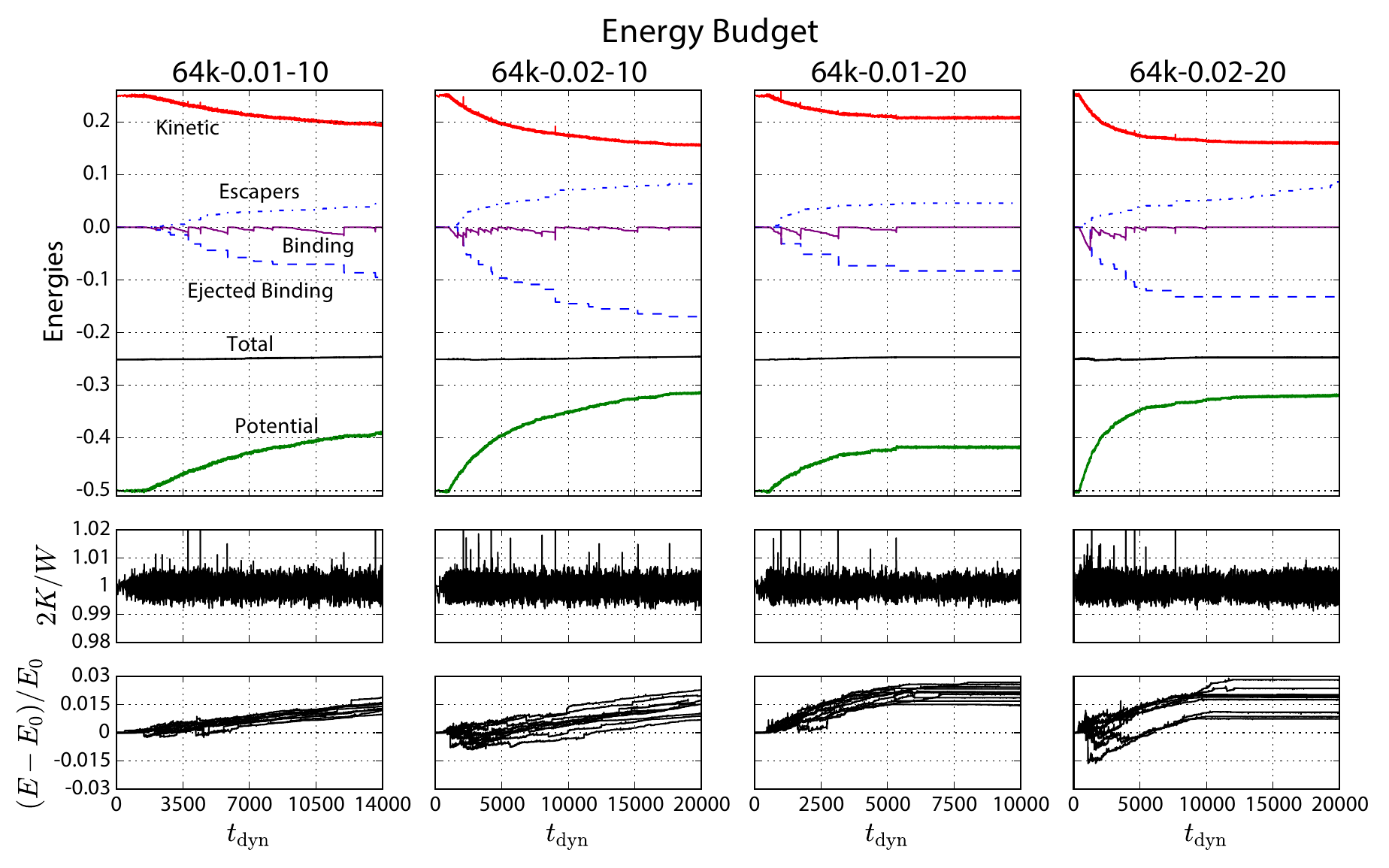}
\label{fig:energy}
\end{figure*}

To check the consistency of the \texttt{RAPID} approach, we examine the energy budget and conservation of each of the runs.  We
quantify the various forms of energy, including the kinetic and potential energy
of all particles, the binding energy of multiples, and the total energy carried
out of the cluster by ejected particles.  The energies are plotted in Figure
11.  In addition, we also consider the virial ratio ($2K/W$)
and the total energy over time for
each system, as a diagnostic of the effectiveness of the method. {The top two
plots show the energy budget and virial ratio for a single representative model,
while the bottom row shows the total energy for all 10 \texttt{RAPID} models.}

The total energy conservation of \texttt{RAPID} can vary over a
single run, and usually lies within 2-3\% of the initial energy for the duration
of the run.  There are two main sources of error which contribute to this
energy flux.   {The first is the difficulty of integrating higher-order
multiples and very close encounters accurately, particularly those that occur in close triple systems.  This issue is not limited to
the Kira integrator, and is one of the well-known issues common to all
collisional dynamics simulations (the so called ``terrible triples''). Although Kira does implement Keplerian regularization for isolated two-body systems and higher-order multiples, we still find that occasionally long-lived triples can induce substantial jumps in energy conservation.}
Furthermore, as the \texttt{CMC} potential is only applied to the center-of-mass
of any multiple systems, any tidal effects upon the multiples from the
background cluster potential are not incorporated correctly.  These integration
errors manifest as discontinuous jumps in the overall energy conservation, which
can be seen in the bottom panels of Fig.\ 10.

The second source of error arises from the integration of orbits in a fixed external
potential.  {This error takes two specific forms.  First, the MC method, as
described above, has an inconsistency in the computation of the potential.  When
a timestep is performed in \texttt{CMC}, the potential is computed first, before the
dynamical encounters take place and the new orbit is calculated.  However, the
new orbit is calculated using the original potential, which does not take in to
account the evolution of the cluster while the particles are dynamically
interacting.  While the work done by neighboring particles is correctly
accounted for, the work done by the change in potential upon each particle is ignored.
}
{To compensate for this energy drift, \texttt{CMC} employs a technique developed by
\cite{Stodoikiewicz1982}, in which the work done by the changing potential is
explicitly added to the kinetic energy of the particle at the end of each
timestep.  This allows energy to be conserved in the MC to 1 part in $10^3$ over
a run \citep{Fregeau2007}.  In \texttt{RAPID}, we self-consistently correct the
velocities of stars controlled by the MC in a similar fashion, but do not account for this energy
drift in the BHs.  We will explore modifications to the external potential (such
as a time-dependent background potential for the Kira integrator) in a future work.
}

Additionally, the $4^{th}$-order Hermite integration scheme, while typical for
collisional stellar dynamics, is known to produce systematic energy errors when
integrating many orbits in a fixed potential \citep[][c.f. Figure 3.21]{Binney2008}.  This
produces a small but systematically positive energy drift while integrating the
BHs over many orbits.    {This is particularly problematic for the Kira 
integrator, which assumes that external potentials are weak perturbations to the 
internal dynamics of the cluster.  While this assumption is valid for a cluster 
evolving in a galactic tidal field, in our current approach, the external 
potential from the MC particles is much stronger than the interparticle forces of the $N$-body integration.  Methods to improve the long-term stability of the $N$-body integration, allowing for many integrations in a fixed potential while still treating close encounters accurately, are currently being investigated}.

This issue may also be exacerbated by the particular combination of the 
dynamical timesteps between MC and $N$-body employed here.  By integrating both 
systems for an identical length of time and combining the results, it is 
entirely possible that \texttt{RAPID} cannot adjust to rapid dynamical changes that occur \emph{during} either the MC or $N$-body integrations.  Investigations into using an adaptive timestep between the two computational domains, similar to the effective operator splitting developed by \cite{2007PASJ...59.1095F} and employed in \cite{2013CoPhC.184..456P}, are currently underway.

Somewhat unexpectedly, Figure 11 shows that this net
energy drift does not depend on the number of BHs present in the cluster, but on
the mass ratio between the individual stars and BHs.  For clusters with a
smaller  $m_{\rm{BH}}/m_{\rm{star}}$, this net energy drift
is slower.  This indicates that for more realistic clusters
containing many BHs and stars of different masses, this energy drift should improve.
This is confirmed by preliminary testing of \texttt{RAPID} on clusters with a
realistic initial mass function.

\section{Conclusion}

In this paper, we described the motivation and development of a new hybrid
technique for dynamically modeling dense star clusters. By combing our Cluster
Monte Carlo (\texttt{CMC}) code with the Kira direct $N$-body
integrator, we are able to combine the speed of the MC approach with
the accuracy of a direct summation.  This hybrid code, the \emph{Rapid and Precisely Integrated Dynamics} (\texttt{RAPID}) Code is designed to
accurately model the non-equilibrium BH dynamics that powers the overall
evolution of GCs and GNs.  Given recent observational detection of BH candidates
in GCs, and the importance of theoretical modeling of GC BHs to X-ray binary
astrophysics \citep{Pooley2003} and gravitational-wave astrophysics \citep{Rodriguez2015a,Antonini2015}, understanding the
dynamics of BHs in clusters is crucial to understanding BH astrophysics.

We found that the hybrid approach is able to replicate both the half-mass radius
and the core radius for several $N$-body models of idealized GCs with a much
greater accuracy than a traditional MC integration.  Unlike a purely
MC approach, \texttt{RAPID} can model the highly non-spherical and
rapidly changing dynamics of the few BHs in the center of the cluster.  This
suggests that the \texttt{RAPID} approach can follow the dynamics of BH systems with
comparable accuracy to a direct $N$-body integration, but with roughly the same
integration time (with in a factor of 2) of an orbit-sampling Monte Carlo approach.

With this technique, it will be possible to explore regions of the GC
parameter space that have remained outside the computational feasibility of
direct $N$-body computations.  In particular, by treating the central BH
subcluster correctly, \texttt{RAPID} can explore regions of the GC and GN
parameter space, including clusters with massive central BHs, that have
previously been unexplored by direct collisional methods.

Two issues remain to be addressed.  First, the Kira $N$-body integrator does not
completely conserve energy in the presence of a large external potential.  This
effect is a well-known drawback of $4^{th}$-order Hermite integrators, will need
to be addressed.  Efforts are currently underway to increase the computational order of the
Hermite predictor-corrector, improving both the accuracy and speed of the
integration \citep[e.g.,][]{Nitadori2008}, and to incorporate a time-dependent
potential in the $N$-body integrator, to account for the work done by the total
cluster potential on the BHs..

Secondly, the regularization of binaries and higher-order multiples in Kira is
based on Keplerian regularization for sufficiently unperturbed systems
\citep[see][]{PortegiesZwart2001a}.  However, this regularization does not include any tidal
effects from the external Monte Carlo potential, which will effect the long-term evolution of binaries retained by the cluster.  Incorporation of these physical effects into
the regularization scheme is currently underway.

Future work will also investigate hardware acceleration of the $N$-body integrator.  The Kira integrator is designed to run on the specialized GRAPE series of hardware, which yields substantial improvements in computational speed.  When combined with the Sapporo GPU/GRAPE library \citep{Gaburov2009}, Kira can be run on modern, distributed GPU systems with comparable performance to the \texttt{NBODY} series of codes \citep{Anders2012}.  Hardware acceleration was not implemented the current \texttt{RAPID} version, since we have not considered systems with sufficiently large numbers of BHs for efficient GPU useage; however, the development of the Sapporo2 library \citep{Bedorf2015} provides efficient GPU saturation for small-$N$ systems.  We will explore the advantages of a \texttt{RAPID} integration with Sapporo2 in a future paper.

\texttt{RAPID} is designed to be a single-purpose code incorporating all the 
necessary physics to model dense star clusters.  However, these ``kitchen-sink'' 
codes, in which many numerical codes are integrated into a single parallel 
infrastructure, are often difficult to extend or modify for different purposes, 
particularly with regard to the shared timestep.    The energy drift noted in 
Section \ref{sec:results:energy} arises from a combination of the 4${^{\rm 
th}}$-order Hermite integrator and the particular combination of the two 
dynamical timesteps.  While the parallel design of \texttt{RAPID} makes it 
difficult to explore variations on this code structure, there do 
exist more modular approaches to computational stellar dynamics that may prove 
helpful.  For example, the Astrophysical Multipurpose Software Environment 
\citep[AMUSE, ][]{2013CoPhC.184..456P} can be used to easily swap different 
$N$-body integrators into a large-scale astrophysics code.  Furthermore, there 
exist methods of combining different dynamical timesteps in a single code 
\citep[e.g., the Bridge approach,][]{2007PASJ...59.1095F}, similar to the operator splitting approach developed by \cite{1991AJ....102.1528W}, that enable large multi-scale simulations to be performed with an adaptive, shared timestep.  Because this leapfrog-esque approach is already implemented in AMUSE (as well as several different $N$-body integrators), we are exploring the possibility of integrating \texttt{RAPID} into AMUSE, allowing for greater precision and flexibility in the $N$-body timestep.


\begin{backmatter}

\section*{Abbreviations}
\begin{itemize}
    \item \textbf{GC} -- Globular Cluster
        \item \textbf{GN} -- Globular Cluster
    \item \textbf{MC} -- Monte Carlo
    \item \textbf{RAPID} -- Rapid and Precisely Integrated Dynamics
    \item \textbf{CMC} -- Cluster Monte Carlo
    \item \textbf{AMUSE} -- Astrophysical Multipurpose Software Environment
    \item \textbf{BH} -- Black Hole
    \item \textbf{SSE} -- Single Stellar Evolution
    \item \textbf{BSE} -- Binary Stellar Evolution

\end{itemize}
\section*{Declarations}
\subsection*{Availability of Data and Materials}
    The authors have elected to not release the \texttt{RAPID} or \texttt{CMC} codes publicly at this time.  Any of the data presented in this paper is available upon request.

\subsection*{Competing interests}
  The authors declare that they have no competing interests.

\subsection*{Funding}
  CR was supported by an
NSF GRFP Fellowship, award DGE-0824162, and is currently supported by the MIT Pappalardo Fellowship in Physics.  This work was supported by NSF Grant AST-1312945 and NASA Grant NNX14AP92G.

\subsection*{Author's contributions}
    The \texttt{RAPID} code was developed in equal parts by CR and BP, with CR developing most of the astrophysical features and BP developing most of the parallel infrastructure; SC, MM, and FR assisted with the former, while AC and W-k L assisted with the later.  CR prepared this manuscript and performed the tests detailed within.

\subsection*{Acknowledgements}
  We thank Eugene Vasiliev, Simon
Portegies Zwart, Vicky Kalogera,
Claude-Andr\'e Faucher-Gigu\`ere, Douglas Heggie,
Philip Breen, and Fabio Antonini for useful discussions.

\bibliographystyle{bmc-mathphys} 






\end{backmatter}

\end{document}